\documentclass[sigconf]{acmart}
\setcopyright{acmcopyright}
\acmYear{2026}

\usepackage{CJKutf8}
\usepackage{fontawesome}
\usepackage{multirow}
\usepackage{makecell}
\usepackage[nointegrals]{wasysym}
\usepackage{booktabs}
\usepackage{pifont}
\usepackage{enumitem}
\usepackage{hyperref}
\usepackage{threeparttable}
\usepackage[table,xcdraw]{xcolor}
\usepackage{microtype}
\usepackage{subfigure}
\usepackage{nicematrix}
\usepackage{graphicx}
\usepackage{etoolbox}
\usepackage{tabularx}
\usepackage{tcolorbox}
\usepackage{makecell}
\usepackage{listings}
\usepackage[misc]{ifsym}
\usepackage[]{footmisc}
\usepackage[skip=2pt,font=small]{caption}
\usepackage{xcolor}
\usepackage{url}
\usepackage{multicol}
\usepackage{stfloats}
\usepackage{pythonhighlight}
\usepackage{adjustbox}
\usepackage{placeins}
\usepackage{appendix}
\usepackage{soul}

\hypersetup{hidelinks}
\AtBeginEnvironment{tablenotes}{\footnotesize}

\makeatletter
\def\UrlAlphabet{%
      \do\a\do\b\do\c\do\d\do\e\do\f\do\g\do\h\do\i\do\j%
      \do\k\do\l\do\m\do\n\do\o\do\p\do\q\do\r\do\s\do\t%
      \do\u\do\v\do\w\do\x\do\y\do\z\do\A\do\B\do\C\do\D%
      \do\E\do\F\do\G\do\H\do\I\do\J\do\K\do\L\do\M\do\N%
      \do\O\do\P\do\Q\do\R\do\S\do\T\do\U\do\V\do\W\do\X%
      \do\Y\do\Z}
\def\UrlDigits{\do\1\do\2\do\3\do\4\do\5\do\6\do\7\do\8\do\9\do\0}
\g@addto@macro{\UrlBreaks}{\UrlOrds}
\g@addto@macro{\UrlBreaks}{\UrlAlphabet}
\g@addto@macro{\UrlBreaks}{\UrlDigits}

\definecolor{codegreen}{rgb}{0,0.6,0}
\definecolor{codegray}{rgb}{0.5,0.5,0.5}
\definecolor{codepurple}{rgb}{0.58,0,0.82}
\definecolor{backcolour}{rgb}{0.95,0.95,0.92}
\definecolor{qinglv}{rgb}{0,0.64,0.59}

\lstdefinestyle{mystyle}{
    backgroundcolor=\color{backcolour},   
    commentstyle=\color{codegreen},
    keywordstyle=\color{magenta},
    numberstyle=\tiny\color{codegray},
    stringstyle=\color{codepurple},
    basicstyle=\ttfamily\footnotesize,
    breakatwhitespace=false,         
    breaklines=true,                 
    captionpos=b,                    
    keepspaces=true,                 
    numbers=left,                    
    numbersep=5pt,                  
    showspaces=false,                
    showstringspaces=false,
    showtabs=false,                  
    tabsize=2
}

\lstdefinelanguage{PHP}{
    keywords={php, if, else, while, do, for, return, function, class, public, private, protected, var, true, false, null, this, new, clone, throw, catch, final, abstract, interface, const, static, namespace, use, global, isset, empty, unset, eval, include, require},
    otherkeywords={<?php, ?>, \$, ::, ->},
    sensitive=true,
    morecomment=[l]{//},
    morecomment=[s]{/*}{*/},
    morestring=[b]",
    morestring=[b]',
}

\lstdefinestyle{prompt}{
    basicstyle=\ttfamily\footnotesize,
    breaklines=true,
    frame=tb,
    framerule=0.5pt,
    backgroundcolor=\colorbox{gray!8},
    xleftmargin=15pt,
    xrightmargin=15pt,
    aboveskip=8pt,
    belowskip=8pt,
    showstringspaces=false
}

\lstdefinestyle{my-python-style}{
  style=pythonhighlight-style,
  frame=none,
  numbers=left,
  numberstyle=\tiny\color{gray},
  escapeinside={(*@}{@*)}
}

\newcommand{\Gu}[1]{{#1}}

\newcommand{\DONE}[1]{\textcolor{blue}{}}

\usepackage{xspace}
\newcommand{\OurTool}{\textsc{Insightor}\xspace}

\newcommand\redsout{\bgroup\markoverwith{\textcolor{red}{\rule[0.5ex]{3pt}{1.5pt}}}\ULon}

\newcommand\toremove[1]{{}}

\AtBeginDocument{%
  }

\setcopyright{acmlicensed}
\copyrightyear{2026}
\acmYear{2026}
\acmDOI{XXXXXXX.XXXXXXX}
\acmConference[ACM CCS '26]{}{Nov 15--19,
  2026}{Hague, Netherlands}
\acmISBN{978-1-4503-XXXX-X/2018/06}

\usepackage{xstring}
\newif\ifsuppressedyear
\makeatletter
\AtBeginDocument{%
  \let\oldbibinfo\bibinfo
  \let\oldnatexlab\natexlab
  \renewcommand{\bibinfo}[2]{%
    \IfStrEq{#1}{year}{%
      \IfStrEq{#2}{[n.\,d.]}%
        {\global\suppressedyeartrue}%
        {\global\suppressedyearfalse\oldbibinfo{#1}{#2}}%
    }{\oldbibinfo{#1}{#2}}%
  }%
  \renewcommand{\natexlab}[1]{%
    \ifsuppressedyear
      \expandafter\@gobbleperiod
    \else
      \oldnatexlab{#1}%
    \fi
  }%
  \def\@gobbleperiod{\global\suppressedyearfalse\@ifnextchar.{\@gobble}{}}%
  \def\NAT@parse@date#1#2#3#4#5#6@@{\def\NAT@year{#1#2#3#4}\def\NAT@exlab{}}%
}
\makeatother

\begin{document}
\begin{CJK*}{UTF8}{gbsn}

\title{Your Space is My Zone: Demystifying the Security Risks of AI-Powered Applications on Pre-Trained Model Hubs}

\author{Yacong Gu}
\affiliation{
  \institution{Tsinghua University; Tsinghua University-QI-ANXIN Group JCN}
  \city{Beijing}
  \country{China}
  }
\email{guyacong@tsinghua.edu.cn}

\author{Lingyun Ying~\textsuperscript{\Letter}}
\affiliation{
  \institution{QI-ANXIN Technology Research Institute}
  \city{Beijing}
  \country{China}}
\email{yinglingyun@qianxin.com}

\author{Zidong Zhang}
\affiliation{
  \institution{QI-ANXIN Technology Research Institute}
  \city{Beijing}
  \country{China}}
\email{zhangzidong@qianxin.com}

\author{Yingyuan Pu}
\affiliation{
  \institution{QI-ANXIN Technology Research Institute}
  \city{Beijing}
  \country{China}}
\email{puyingyuan01@qianxin.com}

\author{Xiaoxue Huang}
\affiliation{
  \institution{QI-ANXIN Technology Research Institute}
  \city{Beijing}
  \country{China}}
\email{huangxiaoxue01@qianxin.com}

\author{Jiawei Zhou}
\affiliation{
  \institution{QI-ANXIN Technology Research Institute}
  \city{Beijing}
  \country{China}}
\email{zhoujiawei01@qianxin.com}

\author{Wenjie Zhu}
\affiliation{
  \institution{QI-ANXIN Technology Research Institute}
  \city{Beijing}
  \country{China}}
\email{zhuwenjie02@qianxin.com}

\author{Donghong Sun}
\affiliation{
  \institution{Tsinghua University}
  \city{Beijing}
  \country{China}
  }
\email{sundonghong@tsinghua.edu.cn}

\author{Haixin Duan}
\affiliation{
  \institution{Tsinghua University}
  \city{Beijing}
  \country{China}
  }
\email{duanhx@tsinghua.edu.cn}

\begin{abstract}
AI-powered Applications (AI-Apps), hosted on platforms such as Hugging Face, are democratizing access to pre-trained models through online inference and fine-tuning services. While lowering AI adoption barriers, these platforms introduce an unexplored attack surface, as AI-Apps are often developed by untrusted parties with weak isolation and misconfigured security settings.
In this paper, we present the first systematic security analysis of AI-Apps across three leading platforms. 
To structure our investigation, we map the AI-App lifecycle to established risk taxonomies (e.g., OWASP), identifying five threat categories and ten attack vectors ranging from generic web flaws to high-impact architectural issues.
Our analysis reveals critical failures including broken access control, insecure resource reuse, insufficient input validation, and sensitive data exposure. Notably, we uncover three novel architectural vulnerabilities inherent to platform design and demonstrate how traditional issues (e.g., world-readable logs) are uniquely amplified in this ecosystem.
To assess real-world impact, we develop an analysis framework \OurTool and apply it to over 970,000 public AI-Apps.
Alarmingly, we find thousands of apps leaking credentials, hundreds containing input injection vulnerabilities that allow arbitrary code execution, and tens harboring embedded backdoors—indicating active exploitation. We have responsibly disclosed all findings to the affected platforms and developers.

\end{abstract}




% Short author string for the running header (9 authors overflow/overlap otherwise)
\renewcommand{\shortauthors}{Gu et al.}

\maketitle

\label{sec:introduction}
\section{Introduction}

With the rapid development of artificial intelligence, numerous Pre-Trained Models (PTMs) have emerged.
Platforms such as Hugging Face now offer AI-powered Applications (AI-Apps, called \emph{Space} on Hugging Face) that enable streamlined model access.
Unlike conventional workflows requiring local deployment and complex environment setup to run models, AI-Apps are cloud services that integrate models with runtime environments, inference APIs, and web interfaces.
This approach simplifies deployment for developers while enabling end-users to perform inference and fine-tuning through browsers or APIs without specialized hardware.
AI-Apps have gained remarkable traction. As of December 2025, Hugging Face hosts over 930,000 AI-Apps, while platforms like Replicate serve major companies including Meta, OpenAI, and Google. For example, Meta's Llama AI-Apps have processed over 600 million inference requests on Replicate~\cite{MetaReplicate}.

Unfortunately, AI-Apps introduce new attack surfaces as they can be created and accessed by less-trusted users. Improper configuration and insufficient isolation may allow adversaries to inject malicious payloads into victim AI-Apps or extract sensitive data from developers and users. 
Recent incidents demonstrate these risks: an official Hugging Face model conversion AI-App contained a vulnerability that allowed attackers to impersonate the conversion bot and manipulate arbitrary AI-Apps~\cite{sestitoHijackingSafetensorsConversion2024}. 
Additionally, security flaws of AI-App Platforms (AAPs) pose significant threats. Replicate once mistakenly assigned shared networks to multiple AI-Apps, enabling attackers to manipulate inference or steal private models through malicious AI-Apps~\cite{SharedNetworkVulnerabilitya}.
Moreover, misuse and misconfiguration of \emph{secrets} (e.g., access tokens for third-party services) on AAPs also introduce the risk of sensitive data exposure. For instance, in 2024, Hugging Face detected unauthorized platform access that compromised multiple users' secrets~\cite{SpaceSecretsSecurity}.

In this paper, we conduct the first systematic study towards understanding security threats in AI-Apps, focusing on three key stakeholders: platforms, developers, and users. 
We conduct a comprehensive analysis of AI-App implementations on three mainstream AAPs (Hugging Face~\cite{SpacesHuggingFace2025}, Replicate~\cite{Replicate}, and ModelScope~\cite{modelscope}), from AI-App creation, user interface, authentication, to resource isolation and sharing.
We find that AI-Apps have some important characteristics that distinguish them from common web applications: they are essentially web servers created by developers and integrated into AAP web pages through iframe embedding. We discover that there are multiple design flaws in AAPs that could potentially cause security risks, enabling attackers to perform various attacks such as code injection and data leakage.

To identify these issues systematically, we map the AI-App lifecycle to established web~\cite{OWASPTop10}~\cite{owasp_llm_top10_2025} and supply-chain~\cite{sok-ssc} risk taxonomies, addressing their dual nature as web applications and composite systems with complex dependencies and cascading trust, prioritizing high-impact and practically exploitable vectors. 
Guided by this methodology, we investigate five types of threats and identify ten potential attack vectors in current AAPs.
Notably, these include three previously unreported, architecture-induced vectors arising directly from platform design choices, as well as known vulnerabilities that manifest with AI-App-specific amplification.
More specifically, \emph{flawed access control} represents a critical vulnerability in multiple AAPs. For example, Hugging Face uses the stateless JWT~\cite{jonesRFC7519JSON} protocol for AI-App authentication, but fails to invalidate tokens when resource ownership changes, enabling attackers to pre-generate tokens and achieve unauthorized access to victim developers' AI-Apps (\emph{Ghost Token Attack}). Additionally, several AAPs misconfigure AI-App runtime logs as \emph{world-readable}, exposing sensitive data from both developers and users (\emph{Log Exfiltration Attack}).
Another critical issue involves \emph{improper resource reuse}, which creates supply chain security risks. When generating access domains for AI-Apps, Hugging Face replaces slashes (/) with hyphens (-) while allowing hyphens in usernames, causing domain name conflicts between different users' AI-Apps (e.g., \texttt{user/app-name} and \texttt{user-app/name} both map to \texttt{user-app-name} subdomain). This enables attackers to conduct malicious code injection against third-party websites embedding AI-Apps (\emph{Identifier Reuse Attack}).
Similarly, \emph{insufficient input validation} affects many AI-Apps that fail to sanitize user input effectively, allowing attackers to execute arbitrary commands through carefully crafted inputs (\emph{Input Injection Attack}).
Furthermore, \emph{sensitive data leakage} occurs through multiple vectors beyond hard-coded secrets in source code. AI-App runtime logs leak substantial sensitive information, including developer-configured \emph{secrets} and user-submitted data such as prompts (\emph{Runtime Log Leakage}).

From our analysis, we find that all three mainstream AAPs exhibit significant security vulnerabilities. 
To evaluate the potential impact of the proposed attacks, we design and implement an analysis framework \OurTool to perform a large-scale measurement across the three AAPs.
\OurTool collects the source code, container images, and metadata of 972,546 public AI-Apps from the three platforms, covering data from September 2024 to December 2025. 
Then it utilizes pre-defined rules to detect potential vulnerabilities, such as ghost token and identifier reuse.
\OurTool further uses data flow analysis to identify input injection vulnerabilities and data leakage threats.

Our measurement demonstrates that security risks in AI-Apps are not merely isolated coding errors but are often amplified by platform-mediated deployment and reuse mechanisms. \OurTool detects widespread vulnerabilities ranging from architectural flaws to active malicious exploitation. 
Notably, we identify a critical disconnect in platform security: \OurTool flagged 936 AI-Apps that followed official secret-handling guidelines yet potentially leaked credentials through misconfigured world-readable logs, with manual sampling confirming many actual leaks. Beyond accidental leakage, we uncover severe injection vectors, with \OurTool identifying 1,442 potential input injection vulnerabilities and 139,475 apps running outdated SDKs (e.g., Gradio) with known RCE flaws.  Furthermore, we find evidence of active exploitation, detecting 27 AI-Apps with embedded backdoors—some persisting for over a year and spreading through platform cloning features.

We evaluated \OurTool's precision and recall on representative samples, confirming it effectively serves as a high-efficiency filter that reduces the search space from 972,546 AI-Apps to manageable candidate sets for manual verification. We have responsibly disclosed identified vulnerabilities to the relevant stakeholders and have received positive feedback, including \$2,369 in bug bounties from Hugging Face.

In summary, the major contributions of this work include:

\begin{itemize}[leftmargin=*]
\item \emph{Systematic Security Analysis.} We present the first comprehensive security study of the AI-App ecosystem. We characterize security-critical mechanisms—including domain-based isolation, stateless token authentication, shared runtime logs, and code propagation—that create a novel attack surface affecting all stakeholders.

\item \emph{Novel Attack Vectors.} We identify ten attack vectors, including three previously unreported, architecture-induced vectors (i.e., Ghost Token, Authentication Bypass, Identifier Reuse) that arise from platform design choices rather than generic web flaws; and we show how known issues manifest with AI-App-specific amplification, e.g., logs leaking prompts/secrets and code propagation through duplication.

\item \emph{Large-Scale Measurement.} 
We design \OurTool, a dataflow analysis framework that automatically scans 972,546 AI-Apps, uncovering widespread potential vulnerabilities including 1,442 input injection flaws, 936 credential leaks despite following official guidelines, 27 backdoors, and 139,475 apps with known RCE flaws.  Responsible disclosure yielded positive feedback and \$2,369 in bug bounties.

\end{itemize}

Our research artifacts, including the source code of \OurTool and all findings, are available at \url{https://anonymous.4open.science/r/AI-App-Demo-FDDE}.

\label{sec:background}
\section{Background}

\subsection{AI-Apps: A Brief Introduction}
AI-Apps are cloud-based services that encapsulate PTMs with runtime environments and inference APIs. They streamline deployment by eliminating dedicated hardware requirements (e.g., GPUs) and complex configuration. Developers transform models into scalable services using platform infrastructure with automatic scaling and parallel inference, while end users access them via web browsers or APIs with minimal setup. Major platforms demonstrate significant adoption: Hugging Face hosts over 930k public AI-Apps, Replicate and ModelScope host 25k and 8k respectively, with participation from companies including Meta, OpenAI, and Google.

\begin{figure*}[t]
\centering
\includegraphics[width=0.85\linewidth]{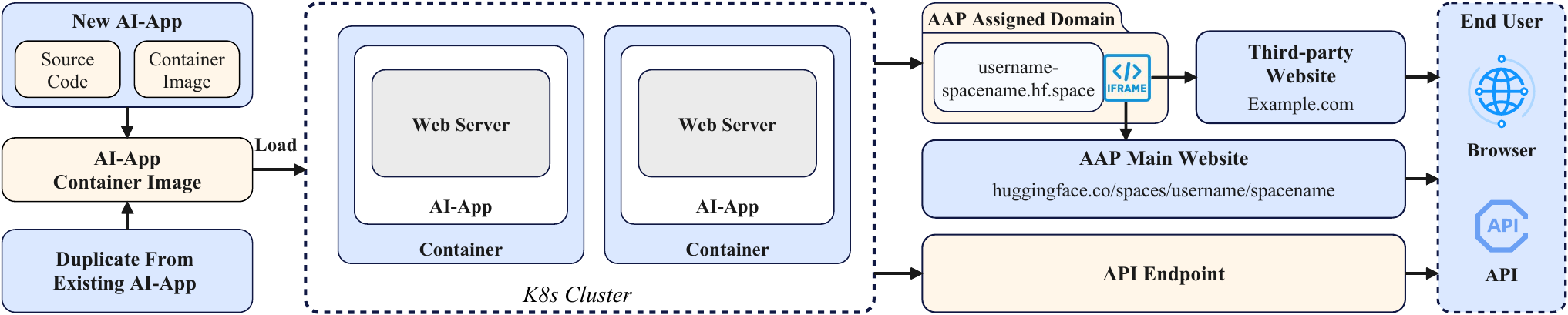}
\caption{The overall architecture of AI-App.}
\label{fig:arch}
\vspace{-2mm}
\end{figure*}

\noindent {\bf AI-Apps Lifecycle.} Figure~\ref{fig:arch} illustrates the typical lifecycle of AI-App. Developers first create an AI-App on an AAP with a globally unique identifier, choosing public or private visibility.
In addition to writing code from scratch, developers can start from official templates based on popular SDKs (e.g., Gradio~\cite{Gradio}) or duplicate existing AI-App source code, similar to the \emph{fork} operation on GitHub. 

Developers then implement the AI-App following the development specifications provided by the AAP. Platforms support two main deployment approaches: some (e.g., Hugging Face, ModelScope) accept source code submissions and automatically build and publish the AI-App, while others (e.g., Replicate) require developers to locally build container images conforming to platform specifications before uploading for deployment.

Upon submission, the AAP allocates resources and launches the AI-App, offering fixed and elastic allocation modes.
Users interact through browser-based web interfaces and API endpoints for service integration. Finally, AAPs provide management functions for modification, updates, deletion, and ownership transfer.

\subsection{Core Concepts of AI-Apps}
\vspace{-1mm}
This section introduces fundamental concepts central to the design, deployment, and utilization of AI-Apps on AAPs.

\noindent{\bf Runtime} refers to the execution environment of an AI-App, which can be either container-based or non-container. The former packages the AI-App into a container image (e.g., Docker) managed by the platform’s orchestration system (e.g., Kubernetes~\cite{ProductionGradeContainerOrchestration}), while the latter is typically utilized for static websites and is limited to applications that do not require GPU resources. Most platforms support \emph{auto-scaling} for container runtimes, enabling developers to configure maximum instances that adjust automatically based on the volume of requests.

\noindent{\bf Graphical User Interface (GUI)} offers a browser-based interface for interacting with an AI-App. It is essentially a developer-implemented web page that takes user input and presents the output. By default, the GUI is rendered on the AAP website, although some platforms also support embedding it into third-party websites~\cite{EmbedYourSpace}.

\noindent{\bf API} support is offered by most AI-Apps, enabling users to integrate AI-Apps into other services. For example, a text classification AI-App can be invoked as part of a larger AI pipeline via API calls.

\noindent{\bf Secrets} are sensitive data required during the execution of an AI-App, such as database credentials or access tokens for third-party services. On most AAPs, the recommended practice is to define secrets as key-value pairs through the platform’s web portal and refer to them by key within the AI-App code, thereby avoiding the exposure of plaintext credentials~\cite{ManageYourSpace}. 

\noindent{\bf Model Training.} AI-Apps that support model training are an important capability on AAPs. Platforms such as Hugging Face offer official training services (i.e., the AI-App \texttt{AutoTrain}~\cite{AutoTrainHuggingFace}), while others provide a framework for developers to build custom training AI-Apps (e.g., Replicate's fine-tune service~\cite{UseModelTraining}). 
Unlike inference applications, training AI-Apps write newly generated models or applications to the user's account upon completion, using platform-managed user credentials to authorize the operation.

\noindent{\bf Billing.} AAPs typically adopt two billing models: charging the developer or the user. 
Whichever person pays the bill, AAPs charge fees based on the time and usage of the computing resource.
To promote sharing, some platforms (e.g., Replicate) implement pricing incentives such as excluding \emph{cold-boot time} from billing for public AI-Apps~\cite{BillingReplicateDocs}, since AI-Apps commonly contain tens of gigabytes of model weights and may incur considerable cold-boot times.

\label{sec:demystified}
\vspace{-1mm}
\section{AI-App Implementation Demystified}
\vspace{-1mm}

Understanding AI-App implementation details is crucial for identifying security vulnerabilities. Since most AAPs provide insufficient documentation, we employ three complementary methods: reviewing AAP documentation, analyzing open-source components (e.g., Hugging Face's \texttt{AutoTrain}~\cite{AutoTrainHuggingFace}, Replicate's \texttt{Cog}~\cite{Cog}), and conducting black-box testing.
For black-box testing, we design a chatbot AI-App that executes user inputs as shell commands and deploy it privately across AAPs. We examine runtime environments (system privileges, environment variables, network isolation), capture network traffic via \texttt{tcpdump}, and create secrets to observe their processing and isolation.
Table~\ref{tab:aap-features} summarizes the key features of examined AAPs.

\vspace{-1mm}
\subsection{AI-App Creation}
\label{subsec:app_creation}
\vspace{-1mm}
Our investigation identifies three distinct approaches for AI-App creation:

\noindent{\bf I. Code Repository-Based Development.} 
Platforms like Hugging Face and ModelScope require developers to commit source code to dedicated repositories. Upon code updates, these platforms automatically trigger build processes that generate container images for deployment.
However, these platforms differ in their security constraints: Hugging Face allows unrestricted base container images, meaning developers can publish AI-Apps with arbitrary—potentially malicious—code, while ModelScope restricts developers to predefined official base images.

\noindent{\bf II. Developer Builds and Pushes AI-Apps.} 
The Replicate platform provides \texttt{Cog}~\cite{Cog}, a Docker-based development tool for building AI-Apps. 
Developers specify dependencies in \texttt{cog.yaml}, implement inference logic in \texttt{predict.py}, and define training procedures in \texttt{train.py}. 
After building the container image locally, developers push it to the platform for deployment.

\noindent{\bf III. Duplicated from Existing AI-Apps.}
All three AAPs support AI-App duplication through different mechanisms. Hugging Face and ModelScope provide \emph{explicit} duplication features, allowing developers to manually fork existing AI-Apps. Replicate implements \emph{implicit} duplication: when developers use AI-Apps with training capabilities, the platform automatically creates new instances under their accounts, replicating the original source code.

\noindent{\bf Billing.} 
The three AAPs adopt different billing strategies. ModelScope employs a developer-paid model, where developers bear usage costs. Replicate follows a user-paid model, charging users directly for AI-App usage. Hugging Face supports both approaches: AI-Apps are billed to developers by default, but the \emph{inference-api} feature enables user authorization and billing~\cite{SignHuggingFace}.
The user-paid model introduces security concerns, as users execute potentially untrusted external code on their computing resources, making them vulnerable to cryptojacking attacks~\cite{FortinetCyrptojacking} where attackers exploit victims' processing power for cryptocurrency mining.

\vspace{-1mm}
\subsection{AI-App User Interface}
\label{subsec:gui}
\vspace{-1mm}
AI-Apps are deployed as web servers providing inference and training services. Platform-specific configurations include: Hugging Face uses the \texttt{app\_port} keyword in \texttt{README.md} files, ModelScope pre-configures ports in base container images, and Replicate's \texttt{Cog} tool injects HTTP servers on port 5000. All three AAPs support browser-based GUI access through two distinct modes:

\noindent {\bf Freestyle Website Mode.} 
Developers maintain full control over the GUI, with content served directly from the AI-App's web server. Hugging Face and ModelScope assign unique subdomains to each AI-App (e.g., Facebook's \texttt{facebook/sapiens-pose} generates \emph{URL\textsubscript{app}}\footnote{https://facebook-sapiens-pose.hf.space/}), while simultaneously creating platform pages (\emph{URL\textsubscript{platform}}\footnote{https://huggingface.co/spaces/facebook/sapiens-pose}) that embed the AI-App via iframes. Similarly, Hugging Face supports embedding AI-Apps in third-party websites via an iframe~\cite{EmbedYourSpace}. 
However, an improperly configured iframe can pose security risks, such as clickjacking and phishing.

\begin{figure}[t]
\centering
\includegraphics[width=0.806\linewidth]{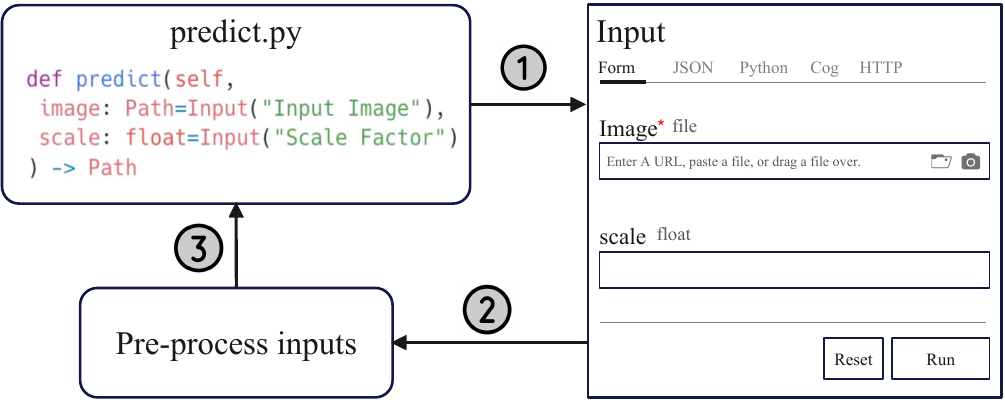}
\caption{Web page generation and input processing on Replicate.}
\label{fig:r8_webpage}
\vspace{-3mm}
\end{figure}

\noindent {\bf AAP-Formatted Web Page.} 
Developers specify input parameters and output formats; AAPs automatically generate web interfaces (step \ding{182} in Figure~\ref{fig:r8_webpage}). For instance, Replicate parses user inputs (step \ding{183}), preprocessing them (e.g., converting files to URLs) and invoking AI-Apps with processed parameters (step \ding{184}).

Beyond web interfaces, all three AAPs support API access and local execution. Replicate provides APIs for all AI-Apps, while most Gradio-based~\cite{Gradio} AI-Apps on Hugging Face and ModelScope also offer API support. For local execution, users can pull container images and run them locally via Docker.

\begin{table*}[htbp]
    \centering
    \small
    \caption{Overview of AI-App implementations. $D$/$U$ in the \textit{Billing} column mean the billing strategies:
$D$ - Developer-Paid, $U$ - User-Paid.}
    \label{tab:aap-features}
    \resizebox{\textwidth}{!}{%
    \begin{tabular}{l c c c c c c c c c c c}
        \toprule
        \multirow{2}{*}{\makecell{\bfseries AAPs}} & \multicolumn{2}{c}{\makecell{\bfseries AI-App Creation}} & \multirow{2}{*}{\makecell{\bfseries Billing}} & \multicolumn{3}{c}{\makecell{\bfseries GUI}} & \multirow{2}{*}{\makecell{\bfseries API}} & \multirow{2}{*}{\makecell{\bfseries Multi-user\\\bfseries Container\\ \bfseries Sharing}} & \multicolumn{2}{c}{\makecell{\bfseries Log Access}} & \multirow{2}{*}{\makecell{\bfseries Secret\\ \bfseries Exposure\\\bfseries in Logs}} \\
        \cmidrule(lr){2-3} \cmidrule(lr){5-7} \cmidrule(lr){10-11}
        & \makecell{\bfseries Full\\ \bfseries Control}
        & \makecell{\bfseries Duplication}
        &
        & \makecell{\bfseries Full\\ \bfseries Control}
        & \makecell{\bfseries Access\\ \bfseries Token}
        & \makecell{\bfseries Embed\\ \bfseries External Website}
        &
        &
        & \makecell{\bfseries Owner}
        & \makecell{\bfseries Any User}
        & \\
        \midrule
        Hugging Face & \checkmark & \checkmark & $D$, $U$ & \checkmark & JWT & \checkmark & \checkmark\rlap{$^*$} & \checkmark & \checkmark & \checkmark & \checkmark\rlap{$^{\dag}$} \\
        Replicate    & \checkmark & \checkmark & $U$   &            &     &            & \checkmark  & \checkmark & \checkmark & \checkmark & \checkmark  \\
        ModelScope   &            & \checkmark& $D$   & \checkmark & \makecell{AAP-defined} & \checkmark & \checkmark\rlap{$^*$} & \checkmark & \checkmark &            & \checkmark  \\
        \bottomrule
    \end{tabular}%
    }
    \begin{tablenotes}\footnotesize
    \item[]{$^*$} AI-Apps built on Gradio can be accessed through API.
    \item[]{$^\dag$} Hugging Face only masks its own platform tokens (strings starting with \texttt{hf\_}), leaving other secrets unprotected.
    \end{tablenotes}
    \vspace{-1mm}
\end{table*}

\vspace{-2mm}
\subsection{Runtime Isolation}
\vspace{-1mm}

\noindent {\bf Inner-App Isolation.}
Our experiments reveal that all three AAPs lack proper isolation between user requests within the same AI-App instance. Different users' requests may share execution contexts, logging systems, and containers. Furthermore, AAPs do not automatically clean temporary files between sessions, relying entirely on developers for cleanup implementation. This design enables persistent attacks where malicious users can compromise shared environments and target subsequent users.

\noindent {\bf Inter-App Isolation.}  
All evaluated AAPs utilize Kubernetes~\cite{ProductionGradeContainerOrchestration} for container orchestration. We assess their implementations against known security risks, including network isolation failures, service account token exposure, and insecure volume mounts. Our analysis does not find such vulnerabilities on any of the three platforms.

\noindent {\bf Network Accessibility.} 
All three AAPs permit AI-Apps to establish outbound connections to arbitrary destinations via standard web ports (80, 443), with Hugging Face additionally supporting port 8080~\cite{SpacesOverview}. This unrestricted external access enables malicious AI-Apps to transmit users' sensitive data or establish reverse shells.

\vspace{-1mm}
\subsection{AI-App Authentication}
\vspace{-1mm}
\label{subsec:app_auth}
This section analyzes authentication mechanisms for AI-App access.

\begin{figure}[t]
\centering
\includegraphics[width=\linewidth]{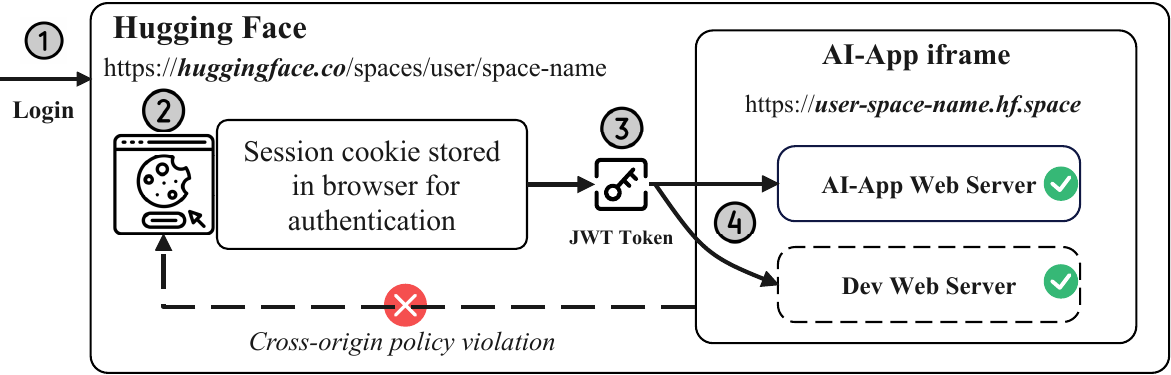}
\caption{Authentication workflow of AI-Apps on Hugging Face.}
\label{fig:hf_jwt}
\vspace{-3mm}
\end{figure}

\noindent {\bf Web Page Access.}
AAPs use session-based authentication for homepage access (\emph{URL\textsubscript{platform}}), storing session keys in cookies upon login (step \ding{182} \ding{183} in Figure~\ref{fig:hf_jwt}). However, accessing AI-Apps directly (\emph{URL\textsubscript{app}}) presents challenges. Hugging Face and ModelScope embed \emph{URL\textsubscript{app}} as iframes within \emph{URL\textsubscript{platform}}, running on separate domains to comply with the same-origin policies, preventing direct cookie inheritance. As a result, the AI-App cannot inherit the user’s authentication state from the \emph{URL\textsubscript{platform}} directly.

Authentication remains essential for AI-Apps, as without it, an attacker could access a private AI-App simply by guessing or obtaining its \emph{URL\textsubscript{app}}. To address this issue, both platforms implement token-based authentication. 
Specifically, Hugging Face employs JWT tokens~\cite{jonesRFC7519JSON} (denoted as \emph{API\textsubscript{jwt}}\footnote{\url{https://huggingface.co/api/spaces/{space\_name}/jwt}}), while ModelScope uses a custom algorithm\footnote{\url{https://modelscope.cn/api/v1/studios/token}}.
When users visit \emph{URL\textsubscript{platform}}, the platform generates an access token (step \ding{184}) appended to \emph{URL\textsubscript{app}} as an HTTP query parameter for user authentication (step \ding{185}). Our analysis reveals that Hugging Face generates JWT tokens for all private AI-Apps, as well as those with \emph{Dev Mode} enabled.

\noindent{\bf Dev Mode Access (Hugging Face).}
Hugging Face's \emph{Dev Mode}~\cite{SpacesDevMode} enables remote development through SSH server installation within containers, supporting VSCode or SSH connections. The browser-based VSCode interface is embedded using an iframe. Like iframe-based AI-Apps, it also uses JWT token authentication. 
\emph{Dev Mode} grants full AI-App control, making unauthorized access particularly risky, as it enables attackers to perform arbitrary code manipulation, backdoor implantation, and secret extraction.

\noindent {\bf Runtime Log Access.}
Hugging Face restricts runtime log access buttons to AI-App developers, which is a sound design decision since logs may contain sensitive data from multiple users and should not be publicly accessible. However, despite this interface-level restriction, both Hugging Face and Replicate configure public AI-Apps' logs to be \emph{world-readable}, creating a significant risk of data leakage. We discuss this issue in detail in the following section.

\vspace{-1mm}
\subsection{AI-App Secrets}
\vspace{-1mm}
AAPs employ two secret management approaches. Hugging Face and ModelScope inject secrets as environment variables. Replicate transmits secrets as plaintext in HTTP request bodies to AI-App web servers, retrieved via \texttt{get\_secret\_value()}~\cite{SecretInputsModels}.

\noindent {\bf Secrets in Logs.}  
Exposing sensitive information in logs is an OWASP top-ten vulnerability~\cite{OWASPTop10}. While services like GitHub Actions implement automatic log scanning to mask registered secrets, none of the examined AAPs provide effective masking, leaving secret values exposed in logs (Table~\ref{tab:aap-features}).
Combined with the runtime log exposure risks identified in Section~\ref{subsubsec:log_exfiltration_attack}, this enables widespread sensitive information leakage.

\vspace{-1mm}
\subsection{Model Training via AI-Apps}
\vspace{-1mm}
\label{subsec:model_train}
Hugging Face and Replicate provide model training services through AI-Apps, creating new models under users' accounts—a process requiring \emph{write} permissions. 
Hugging Face employs \texttt{AutoTrain}~\cite{AutoTrainHuggingFace}, an official training AI-App that users must first duplicate in their accounts.
During training, \texttt{AutoTrain} requests OAuth write access to create and upload the trained model.

Replicate requires developers to implement a \texttt{train()} function, which enables training functionality. Users initiate training via website or API, specifying a destination AI-App. Replicate creates it internally without exposing tokens to the original AI-App, mitigating direct token leakage.
However, the primary security risk lies in Replicate's AI-App replication process. The newly generated AI-App inherits the original's code, receiving trained weights via \texttt{COG\_WEIGHTS} environment variable. 
This enables code propagation attacks where malicious code transfers to newly created AI-Apps.

\label{sec:threats}
\section{AI-App Threats}
\label{sec:aip_threats}

\subsection{Threat Model}
We consider two deployment scenarios for AI-Apps: (1) public deployment accessible to global users and (2) private deployment within organizations with restricted access.
We assume that AAPs are trustworthy providers and that underlying communication channels (e.g., HTTPS) remain secure.

\noindent{\bf Scope and Threat Identification.}
This paper investigates security risks in the AI-App platform ecosystem. We focus on vulnerabilities arising from the shared infrastructure and integration logic, rather than the mathematical properties of the models. Specifically, our analysis covers: 
(1) \emph{Runtime Environments,} covering container lifecycle, image building, and isolation boundaries;
(2) \emph{Interaction Interfaces,} including iframe embeddings, frontend wrappers, and API endpoints; and
(3) \emph{Platform Services,} such as token-based authentication, logging, and storage.
Consequently, model-centric attacks such as \emph{Prompt Injection} are out of scope, as they target model properties rather than the deployment infrastructure.

To systematically identify threats within this scope, we structured our analysis by: first, enumerating security-critical interfaces (e.g., secret handling, token propagation); and second, prioritizing high-impact vectors (e.g., data theft, RCE). Finally, to ground our findings in established security practices, we map the identified attack vectors to standard risk frameworks, including the OWASP Top 10 for Web and LLMs~\cite{owasp_llm_top10_2025} and supply-chain risk taxonomies~\cite{sok-ssc}. A detailed mapping of these vectors is provided in Table~\ref{tab:threat_mapping} in Appendix.

\noindent{\bf Adversaries.} Under these assumptions, we identify two types of adversary: 
(1) \emph{Unethical AI-App developers} may publish malicious applications to exploit users. These AI-Apps can steal sensitive data, conduct phishing attacks, or abuse computing resources for unauthorized activities (e.g., cryptocurrency mining). Attackers may embed malicious code in container images while maintaining clean source code on AAPs, or exploit distribution vulnerabilities to inject malicious applications into third-party websites.
(2) \emph{Unethical AI-App users} may exploit vulnerabilities in legitimate AI-Apps or AAP infrastructure to exfiltrate data or install backdoors. These attackers typically possess basic reverse engineering skills to extract source code, identify common vulnerabilities (e.g., unsanitized inputs), and analyze logs for secrets or user-submitted sensitive data such as prompts. Such attacks require minimal technical effort and cost, particularly in public AI-Apps where developers cover usage costs, making them both practical and scalable.

Table~\ref{tab:threat-overview} summarizes the investigated AAPs and their associated threats that we find.

\begin{table}[t]
    \centering
    \small
    \caption{Overview of potential threats on each AI-App platform. The \ding{51} means vulnerable. HF - Hugging Face, MS - ModelScope.}
    \label{tab:threat-overview}
    \begin{tabular}{lccc}
        \toprule
        \textbf{Threats} & \textbf{HF} & \textbf{Replicate} & \textbf{MS} \\
        \midrule
        {A1: Log Exfiltration} & \checkmark   & \checkmark   &              \\
        \rowcolor[HTML]{EFEFEF}
        {A2: Ghost Token} & \checkmark   &              &              \\
        {A3: Auth Bypass} & \checkmark   & \checkmark   & \checkmark   \\
        \rowcolor[HTML]{EFEFEF}
        {A4: Over-privileged iframe} & \checkmark   &              &              \\
        {R1: Identifier Reuse} & \checkmark   &              &              \\
        \rowcolor[HTML]{EFEFEF}
        {R2: Code Poisoning} & \checkmark   & \checkmark   & \checkmark   \\
        {V1: Input Injection} & \checkmark   & \checkmark   & \checkmark   \\
        \rowcolor[HTML]{EFEFEF}
        {L1: Runtime Log Leakage} & \checkmark   & \checkmark   &              \\
        {L2: Files Leakage} & \checkmark   & \checkmark   & \checkmark   \\
        \rowcolor[HTML]{EFEFEF}
        {P1: Cryptojacking} & \checkmark\rlap{$^*$}  & \checkmark   &              \\
        \bottomrule
    \end{tabular}%
    \begin{tablenotes}\footnotesize
    \item[] $^*$ AI-Apps that employ user-pays models are vulnerable.
    \end{tablenotes}
\end{table}

\vspace{-2mm}
\subsection{Flawed Access Control}
\vspace{-1mm}
Access control defects are among the most common and dangerous security vulnerabilities. Our research reveals that all three AAPs suffer from improper access control design, which can lead to unauthorized access and data leakage.

\subsubsection{Log Exfiltration Attack (A1).}
\label{subsubsec:log_exfiltration_attack}
Runtime logs often contain sensitive data when developers inadvertently logging confidential information. The 2023 TravisCI incident illustrates this risk: 4.7 million exposed logs leaked over 73,000 credentials~\cite{DoesTravisCI}.

Our analysis reveals that Hugging Face and Replicate misconfigure public AI-Apps' runtime logs as \emph{world-readable}, making them vulnerable to log exfiltration attacks. 
Despite hiding the log access interface (i.e., a button on the AI-App's webpage), Hugging Face's \emph{API\textsubscript{jwt}} remains publicly accessible and can issue JWT tokens for any public AI-App, granting complete log access with the platform's log API\footnote{\url{https://api.hf.space/v1/{space\_name}/logs/run}}.
This effectively allows any Hugging Face user to read the runtime logs of any public AI-App.
Similarly, Replicate exhibits the same vulnerability but with limited scope/impact. Specifically, Replicate assigns each task (inference or model training) a random unique identifier consisting of 26 characters.
Although possessing an identifier enables log retrieval, the platform lacks identifier enumeration methods. This limits the attack surface to logs explicitly shared by developers, as brute-force identifier guessing remains computationally infeasible.
In contrast, ModelScope restricts log access to administrative users only, eliminating this attack vector.

\vspace{-1mm}
\subsubsection{Ghost Token Attack (A2).}

Hugging Face allows reusing deleted account names, creating identifier (i.e., \texttt{user/spacename}) collisions.
Moreover, Hugging Face's \emph{API\textsubscript{jwt}} is publicly accessible and allows users to generate JWT tokens for specified AI-Apps.
Thus, combined with JWT's \emph{stateless} nature, previously issued tokens remain valid after resource ownership changes.

We find Hugging Face vulnerable to this attack. Attackers can predict future AI-App identifiers (e.g., by targeting organizations active on GitHub but absent from Hugging Face). They register the predicted username, create a private AI-App, generate a JWT token (valid for 24 hours), and then immediately delete both resources. Due to JWT's stateless nature, the token remains valid despite resource deletion.
When victims later register the same username and (re)create the predicted AI-App, the attackers' token retains access even if the victim AI-App is private. This attack relies on two factors: (1) \emph{accurate identifier prediction}—enabled by username overlap between GitHub and Hugging Face (Section~\ref{subsubsec:ghost_token_result}); (2) \emph{token validity maintenance}—achieved through repeated resource creation/deletion cycles using free accounts at minimal cost.

\vspace{-1mm}
\subsubsection{Authentication Bypass Attack (A3).}

We identify AI-Apps implementing flawed access control through hard-coded password verification across two scenarios.
In the first scenario, developers embed password verification logic in publicly visible AI-Apps to restrict unauthorized access. For example, ardianfe/stable-audio-prod on Replicate verifies whether the user-supplied \texttt{key} parameter matches a hard-coded string (line 4 in Listing~\ref{lst:password-demo}), terminating execution upon failure. One likely reason for this approach is Replicate’s billing model, which eliminates cold-boot charges for public AI-Apps~\cite{BillingReplicateDocs}, while attempting to maintain access control through rudimentary authentication.
The second scenario involves integrating admin panels into AI-Apps, where password entry grants access to AI-App management interfaces. While such panels may facilitate administration, proper authentication should rely on \emph{secrets} rather than hard-coded passwords.

Both scenarios expose critical security vulnerabilities: since passwords are hard-coded in source code within container images, attackers can extract and examine these images to trivially retrieve credentials and bypass authentication.

\lstset{
numbers=left,
numbersep=-0.8mm}
\begin{python}[style=my-python-style]
 def predict( # .... other parameters
     key: str = Input(description="key password") 
   ) -> dict:
     if key != "#p'7y9EqH4>el;VZ'Ok*RK#Jt,/old":
       raise ValueError("Invalid key password")
\end{python}
\begin{lstlisting}[frame=none,caption={Checking for hard-coded passwords in AI-App \texttt{ardianfe/stable-audio-prod} on Replicate.},captionpos=b,label=lst:password-demo]
\end{lstlisting}

\vspace{-1mm}
\subsubsection{Over-privileged iframe (\emph{A4}).}

Hugging Face and ModelScope embed AI-Apps using \texttt{iframe} elements with overly broad permissions. Our analysis reveals that Hugging Face grants each AI-App iframe up to 27 permissions by default\footnote{accelerometer, ambient-light-sensor, autoplay, battery, camera, clipboard-read, clipboard-write, display-capture, document-domain, encrypted-media, fullscreen, geolocation, gyroscope, layout-animations, legacy-image-formats, magnetometer, microphone, midi, oversized-images, payment, picture-in-picture, publickey-credentials-get, sync-xhr, usb, vr, wake-lock, xr-spatial-tracking}, including access to camera and microphone.

This over-permissioned design creates a privilege inheritance vulnerability. 
When users grant permissions to legitimate AI-Apps, subsequently loaded malicious applications inherit these permissions without additional user consent. Consequently, attackers can silently access sensitive resources---such as camera feeds or clipboard data---by bypassing browser security prompts.

\subsection{Improper Resource Reuse}
AI-Apps introduce resource reuse vulnerabilities that attackers exploit via identifier mapping and code inheritance to compromise applications and spread malicious content.

\subsubsection{Identifier Reuse Attack (R1).}
\label{subsubsec:id_reuse_attack}

Improper identifier reuse poses classic supply chain security risks~\cite{gu2023investigating}, and AI-Apps are not exempt from this threat. 
We identify a domain takeover vulnerability on Hugging Face that enables silent malicious code injection through orphaned iframe links.

Hugging Face generates subdomains for AI-Apps by replacing slashes with hyphens in their identifiers (e.g., \texttt{user/space-name} → \texttt{user-space-name.hf.space}). Since usernames can contain hyphens, distinct AI-Apps (\texttt{user/space-name} and \texttt{user-space/name}) may collide on the same subdomain.
Hugging Face resolves conflicts by appending random suffixes to new subdomains.

\begin{figure}[t]
\centering
\includegraphics[width=0.95\linewidth]{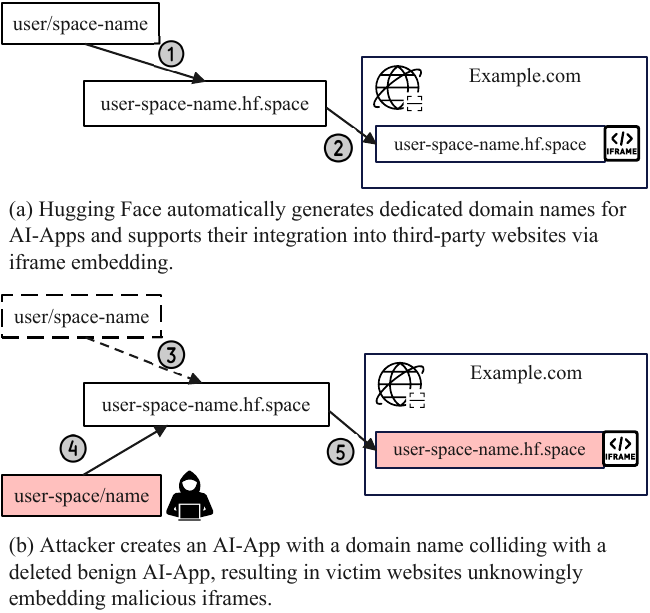}
\caption{Overview of Identifier Reuse Attack (\emph{R1}).}
\label{fig:id_reuse_attack}
\vspace{-2mm}
\end{figure}

This creates an attack vector: attackers pre-register accounts (e.g., \texttt{user-space}) and monitor for target AI-App deletions. When \texttt{user/space-name} is deleted, iframes embedded in third-party websites become dangling resources pointing to unassigned subdomains. Attackers then create \texttt{user-space/name}, thereby reclaiming the original subdomain \texttt{user-space-name.hf.space}. 
Notably, Hugging Face does not append a random suffix in this case, as no conflicting AI-Apps exist on the platform at this point.
Consequently, previously benign iframes now load malicious content from the trusted Hugging Face domain (Figure~\ref{fig:id_reuse_attack}).

This attack is difficult to detect, as iframes appear legitimate from official Hugging Face domains without requiring code modifications by website owners.
Takeovers can persist undetected, enabling sustained phishing, credential harvesting, and targeted attacks.

\subsubsection{AI-App Poisoning Attack (R2).}
AI-Apps can reuse code through two primary mechanisms identified in Section~\ref{subsec:app_creation}.
First, developers may explicitly duplicate existing AI-Apps. Second, our investigation reveals that Replicate automatically generates new AI-Apps inheriting source code when users perform model training, without explicitly notifying developers of this code replication.

This behavior creates opportunities for malicious exploitation. An unethical developer could publish an AI-App containing malicious code (e.g., backdoors) and wait for victims to copy or unknowingly trigger code duplication. Such attacks can result in severe consequences, as poisoned AI-Apps may leak sensitive data or provide unauthorized access to private infrastructure when deployed internally. We find that all three AAPs are vulnerable to this threat.

\subsection{Insufficient Input Validation}
As web-based applications, AI-Apps face similar input injection risks to traditional web services.

\subsubsection{Input Injection Attack (V1).}
All three AAPs are vulnerable to input injection attacks through two primary vectors.
First, developers may implement insecure code that inadequately sanitizes user input, enabling malicious command injection. 
Second, AI-Apps may rely on vulnerable third-party SDKs that remain unpatched. 
For example, Gradio---a widely-used front-end library for AI-Apps---has suffered multiple code injection vulnerabilities~\cite{NVDCve20236572}\cite{NVDCVE20241540}\cite{NVDCVE20241728}\cite{NVDCVE202439236}\cite{NVDCVE202447867}, compromising dependent AI-Apps.

Input injection poses serious threats to both developers and end users. 
Attackers can execute arbitrary shell commands to extract sensitive data (e.g., \emph{secrets} in environment variables) or inject persistent malicious code that compromises subsequent users.
Listing~\ref{lst:input-injection-demo} demonstrates this vulnerability in an AI-App that directly concatenates user-supplied parameters into shell commands (line 7), enabling arbitrary command execution.

\lstset{
numbers=left,
numbersep=-0.8mm}
\begin{python}[style=my-python-style]
 def predict(
    # other parameters
    speaker: str = Input(description="Original 
      speaker audio url...")   (*@{\fontsize{8.5}{10}\selectfont\color{red}{\;\faBomb} User Input}@*)
  ) -> Any:
    .....
    os.system(f"ffmpeg -i {speaker} -af 
      {filter}{trim_silence} -y {speaker_wav}")
\end{python}
\begin{lstlisting}[frame=none,caption={Code injection vulnerability in the AI-App \texttt{codehappynice/voicegenerator} on Replicate.},captionpos=b,label=lst:input-injection-demo]
\end{lstlisting}

\subsection{Sensitive Data Leakage}
We find all three AAPs suffer from serious data exposure issues, affecting numerous AI-Apps across platforms.

\subsubsection{Runtime Log Leakage (L1).}

Writing sensitive data into log files is an OWASP Top 10 security vulnerability~\cite{OWASPTop10}.
In AI-App ecosystem, this risk extends to critical assets like proprietary prompts, mapping directly to \emph{OWASP LLM02: Sensitive Information Disclosure} and \emph{LLM07: System Prompt Leakage}~\cite{owasp_llm_top10_2025}.
These risks are amplified by the platform misconfigurations identified in Section~\ref{subsubsec:log_exfiltration_attack}: both Hugging Face and Replicate expose runtime logs to public access, allowing any user—or attacker—to retrieve them for public AI-Apps.

The multi-user nature of AI-Apps makes runtime logs particularly vulnerable to data exposure. We identify three leaked data categories: (1) \emph{Secrets} unintentionally printed during execution, (2) JWT tokens provided by Hugging Face for authentication, and (3) sensitive user inputs/outputs, including LLM prompts and other confidential data.

\vspace{-1mm}
\subsubsection{Container Files Leakage (L2).}
Container images in AI-Apps are susceptible to sensitive data exposure through two primary vectors. 
First, AI-App source code may contain hard-coded credentials, API keys, or tokens.
Second, AI-Apps frequently employ custom container images that contain preinstalled tools, scripts, and configuration files. Without comprehensive auditing, these images may inadvertently include private credentials.
Our analysis demonstrates that this vulnerability affects AI-Apps on all three platforms via source code and image layers.

\subsection{Cryptojacking}
\vspace{-1mm}
We finally examine an attack vector that exploits the unique characteristics of AI-Apps.

\subsubsection{Cryptojacking Attack (P1).}

AI-Apps are attractive targets for cryptojacking due to their heavy reliance on high-performance GPU resources. This threat is particularly acute when users execute code from untrusted sources in two contexts: (1) pulling AI-Apps for local execution, or (2) running them on AAPs under a user-pays model.
In both scenarios, attackers can embed mining algorithms into seemingly benign AI-Apps. These compromised applications deliver expected AI outputs while covertly hijacking the victim's GPU resources for mining operations.

\section{Measurement Methodology}
\label{sec:measurement_methodology}

To investigate the current security status of AI-Apps, we conduct a large-scale measurement on three mainstream AAPs. 

\vspace{-1mm}
\subsection{Data Collection}
We collect all available AI-Apps hosted on the three AAPs, including their code repositories, container images, and metadata (e.g., identifier, developer name, and SDK version), if available. Since each platform exposes different interfaces, we tailor our approach accordingly.

\noindent{\bf Hugging Face.} We query the official API~\cite{HubAPIEndpoints} to obtain public AI-Apps and use the platform's CLI tool~\cite{CommandLineInterface} to download source code and Docker images~\cite{RunDocker}. We collect metadata for 938,602 AI-Apps, with 713,613 (76.03\%) that have downloadable container images. The remaining AI-Apps primarily use \emph{static} runtimes that do not rely on containers.

\noindent{\bf Replicate.}
Replicate's official API~\cite{HTTPAPIReplicate} provides only 5,325 AI-Apps, as it excludes those without the \emph{Example Task}, thereby omitting a large portion of public applications.
To expand our dataset, we employ two strategies: (1) extracting developer names to query all AI-Apps under each account, and (2) enumerating top 5K GitHub organizations and users~\cite{GitstarRankingTop} as candidate usernames. 
We collect 25,340 AI-Apps with 19,412 downloadable container images. The remaining AI-Apps lack downloadable images because some have only registered AI-App pages without releasing any versions, while others are not designed for local execution (e.g., Anthropic's Claude models\footnote{https://replicate.com/anthropic/claude-3.7-sonnet} are proprietary). Unlike other platforms, Replicate does not maintain public source code repositories.

\noindent{\bf ModelScope.} Following a similar approach to Hugging Face, we download the source code repositories and container images. We collect 8,604 AI-Apps and 12 official template container images.

Overall, the total number of AI-Apps that we collect across the target platforms is as follows: Hugging Face (938,602), Replicate (25,340), and ModelScope (8,604).

\subsection{Basic Threat Analysis}
\OurTool identifies potentially vulnerable AI-Apps using heuristic rules derived from our threat models and platform-specific observations. We classify vulnerabilities as follows.

\noindent{\bf Platform-Wide Design Flaws.}
Certain threats stem from systemic platform-level issues. \emph{Ghost Token Attack (A2)} and \emph{Over-privileged iframe (A4)} originate from flawed permission or token management policies, affecting all AI-Apps on the respective platforms.

\noindent{\bf Deployment-Triggered Vulnerabilities.}
Several threats arise from specific deployment configurations. \emph{Log Exfiltration Attack (A1)} affects all public Hugging Face AI-Apps and Replicate AI-Apps that have enabled the Example Task feature. Similarly, \emph{Cryptojacking Attack (P1)} targets AI-Apps operating under user-pays billing models, as well as any AI-App pulled by users for local execution.

\noindent{\bf Developer-Controlled Risk Factors.}
Some threats result from developer implementation choices. \emph{Authentication Bypass Attack (A3)} occurs when user verification relies on hard-coded passwords. \emph{Identifier Reuse Attack (R1)} affects Hugging Face AI-Apps whose identifiers contain hyphens and overlap with unregistered usernames, creating subdomain ambiguity. \emph{AI-App Poisoning Attack (R2)} targets AI-Apps created through code duplication or model training workflows.

\noindent{\bf Complex Vulnerability Classes.}
\emph{Input Injection Attack (V1)} and \emph{Sensitive Data Leakage} require deeper analysis due to their nuanced characteristics. We detail their detection methodology and findings in subsequent sections.

\begin{table}%
\centering
\small
\caption{Sources and sinks used in data flow analysis.}
\label{tab:taint}
\begin{tabular}{lll}
\toprule
\bfseries Categories & \bfseries Sources & \bfseries Sinks \\
\midrule

Input Injection &
\makecell[l]{%
predict(:str) in Replicate \\
train(:str) in Replicate \\
Textbox() in Gradio \\
text\_input() in Streamlit  \\
text\_area() in Streamlit
} &
\makecell[l]{%
os.system \\
os.popen \\
eval \\
exec \\
subprocess.*%
} \\
\midrule

User Input Leakage &
\makecell[l]{%
predict(:str) in Replicate \\
train(:str) in Replicate \\
Textbox() in Gradio \\
text\_input() in Streamlit  \\
text\_area() in Streamlit
} &
\makecell[l]{%
print \\
logger.* \\
logging.* \\
log.*%
} \\
\midrule

Secret Leakage &
\makecell[l]{%
os.environ.get() \\
os.getenv() \\
os.environb.get()%
} &
\makecell[l]{%
print \\
logger.* \\
logging.* \\
log.*%
} \\
\bottomrule
\end{tabular}%
\vspace{-2mm}
\end{table}

\vspace{-1mm}
\subsection{Input Injection Vulnerability Analysis}
\label{subsec:input_injection}

\OurTool analyzes two categories of input injection risks from different sources.

\noindent{\bf Vulnerabilities in AI-App Code.}
This category stems from improper input sanitization by developers. \OurTool performs data flow analysis on Python-based AI-Apps (the most prevalent type) to trace user inputs (i.e., sources) to dangerous execution functions (i.e., sinks), such as \texttt{os.system()} and \texttt{eval()}.

Source identification requires going beyond standard HTTP parameters, as AI-Apps route inputs through high-level SDK components or predefined template functions that generic taint rules miss.
Specifically, we handle two types of AI-Apps: (1) \textit{Template-based AI-Apps} (e.g., those on Replicate) implement predefined functions like \texttt{predict()} and \texttt{train()}. \OurTool treats string-type input parameters in these functions as sources.
(2) For \textit{Non-template AI-Apps} such as those on Hugging Face and ModelScope, we focus on those built with the two most common SDKs: Gradio~\cite{Gradio} and Streamlit~\cite{StreamlitFasterWay}. These account for a substantial portion of AI-Apps on both platforms (48.61\% on Hugging Face and 69.08\% on ModelScope). We extract input-receiving code patterns from these SDKs as sources, summarized in Table~\ref{tab:taint}.
We implement a CodeQL~\cite{CodeQL}-based data flow analyzer to detect potential vulnerabilities, then manually verify flagged cases.

\noindent{\bf Vulnerable Third-Party SDKs.}
This category arises from security flaws in third-party SDKs rather than developer code. We identify Gradio versions used in AI-Apps on Hugging Face and ModelScope, then check them against the OSV vulnerability database~\cite{OSVOpenSource} for known exposures.

\vspace{-1mm}
\subsection{Data Leakage Analysis}
\label{subsec:data_leakge}

We analyze three categories of sensitive data leakage, each requiring distinct detection approaches.

\noindent{\bf Hard-coded Token Leakage.}
We extract all text files from container images and code repositories, then apply \textsc{KeySentinel}~\cite{zhou2025hey}, a state-of-the-art open-source secret scanning tool, to identify hard-coded token leaks.

\noindent{\bf JWT Token Leakage.}
JWT tokens are primarily leaked through web server frameworks that automatically log HTTP requests. We examine popular frameworks (Flask~\cite{Flask}, FastAPI~\cite{FastAPI}) to identify logging behaviors that capture JWT tokens in runtime logs.

\noindent{\bf Code-guided Log Leakage.}
Log files may expose sensitive data that conventional scanners miss, including non-standard formats of \emph{secrets} (e.g., webhook URLs), and valuable user inputs like LLM prompts that lack typical credential patterns.

We formulate this as a data flow analysis problem from sensitive sources to log sinks. Sources include (1) developer-defined \emph{secrets} accessed via environment variables, and (2) user input variables (Table~\ref{tab:taint}). Since Hugging Face publicly exposes secret keys (as a reminder to developers which secret should be set when duplicating AI-Apps to ensure proper functionality), we treat these variables as sources. Sinks comprise common logging functions. We perform a data flow analysis based on CodeQL to detect leakage, followed by manual validation.

\vspace{-1mm}
\subsection{Limitations}
Our study has several limitations. First, we focus primarily on container-based AI-Apps. Non-container AI-Apps, which comprise 22.95\% of Hugging Face AI-Apps and 19.22\% of ModelScope AI-Apps (with none on Replicate), pose relatively lower security risks due to their limited customization capabilities. Therefore, we do not prioritize their investigation.
Second, \OurTool is designed as a \textit{scalable screening pipeline} rather than a fully precise verifier. Consequently, it is subject to the precision and recall trade-offs of static analysis; we explicitly evaluate its False Positive/False Negative (FP/FN) in Section~\ref{subsec:eval} to validate its effectiveness.
Third, while our analysis methods apply to private AI-Apps, our measurement is constrained to publicly accessible resources, and the results do not cover security risks in private AI-Apps, particularly those deployed within organizations.

\section{Measurement Results}

We adopt \OurTool to conduct the evaluation on three mainstream AAPs and identify that numerous AI-Apps are vulnerable to the threats mentioned in this paper. 

\subsection{Flawed Access Control}

\subsubsection{Log Exfiltration Attack (A1).}
Our analysis reveals that runtime log files expose a large volume of secrets and user inputs. 
Detailed findings are presented in Section~\ref{subsubsec:secret_leakage}.

\subsubsection{Ghost Token Attack (A2).}
\label{subsubsec:ghost_token_result}
The ghost token attack requires attackers to accurately predict future AI-App identifiers. To assess feasibility, we examine whether attackers could leverage public GitHub data to anticipate organizations usernames on Hugging Face.
Among 490,960 Hugging Face users with at least one AI-App, 8,462 (1.72\%) are organizational accounts, of which 4,329 (51.16\%) have usernames matching existing GitHub organizations.

We then examine the similarity of the names between AI-Apps and GitHub repositories within the same organizations. 
Among 10,548 AI-Apps from the 4,329 overlapping accounts, 515 (4.88\%) have names that exactly matched repository names in their corresponding GitHub organizations. For instance, \emph{Mozilla Foundation}'s \emph{mozilla-ai} account exists on both platforms, with 7 of 10 AI-Apps sharing names with GitHub repositories.
This substantial overlap demonstrates that the GitHub data can effectively predict Hugging Face AI-App names, making this attack feasible in practice.

\subsubsection{Authentication Bypass Attack (\emph{A3})}

We identify 65 AI-Apps on Replicate and Hugging Face that use hard-coded passwords for authentication (Table~\ref{tab:auth_byss_result} shows examples).
We conduct a case study on \texttt{Metamajorwsu/WSUMajorExplorer}, a recommendation system developed by Wichita State University based on real student data to help students choose their majors. It integrates an admin panel that records student submissions and authenticates with a hard-coded password.
Attackers can extract admin passwords from source code to access students' private data.
Notably, we have reported this issue to the developer (Section~\ref{subsec:disclosure}).

\begin{table}%
    \centering
    \small
    \caption{AI-Apps with hard-coded authentication passwords. \emph{Entry}: password required for usage; 
    \emph{Admin}: password grants management panel access.}
    \label{tab:auth_byss_result}
    \begin{tabular}{lll}
        \toprule
        \textbf{AI-Apps} & \textbf{AAPs} & \textbf{Types} \\
        \midrule
        \texttt{ardianfe/stable-audio-prod}    & Replicate     & Entry \\
        \texttt{krtr/chatbot\_ventas\_isuzu}    & Hugging Face  & Entry \\
        \texttt{Metamajorwsu/WSUMajorExplorer} & Hugging Face  & Admin \\
        \texttt{Aleksmorshen/Fullsumtestbase}  & Hugging Face  & Admin \\
        \texttt{Iammcqwory/Gamuu}              & Hugging Face  & Admin \\
        \bottomrule
    \end{tabular}%
\end{table}

\subsubsection{Over-privileged iframe (\emph{A4})}
\OurTool scans source code to identify explicit permission requests. On Hugging Face, 21,444 AI-Apps (2.28\%) request at least one browser permission. The most frequently requested ones are \texttt{clipboard-write} (14,742; 1.57\%), \texttt{autoplay} (4,930; 0.53\%), \texttt{camera} (4,169; 0.44\%), \texttt{encrypted-media} (3,687; 0.39\%),  and \texttt{microphone} (3,525; 0.38\%).

While permission requests are not inherently malicious, widespread use of sensitive permissions creates significant security risks. Permissions granted to trusted AI-Apps persist across sessions, allowing embedded AI-Apps to inherit them without user consent.
For instance, after granting camera access to a benign portrait-generation AI-App, a subsequent malicious image-enhancement AI-App can access the camera without prompts through permissive iframe configurations, enabling unauthorized data capture.

\subsection{Improper Resource Reuse}

\subsubsection{Identifier Reuse Attack (R1).}
\label{subsubsec:id_reuse_attack_result}

As discussed in Section~\ref{subsubsec:id_reuse_attack}, an attacker can exploit deleted AI-Apps with hyphen-containing identifiers that were previously embedded in external websites. We evaluate the feasibility and impact of this attack from three perspectives.

\noindent{\bf Vulnerable AI-Apps.} We identify AI-Apps with hyphen-containing identifiers whose collision-prone usernames remain unregistered. Among 938,602 analyzed Hugging Face AI-Apps, 439,887 (47.03\%) contain hyphens. Of these, 438,967 (99.79\%) have at least one unregistered username that attackers could exploit after deletion.

\noindent{\bf Deletion Frequency.} During 30 days of monitoring Hugging Face (2025/04/16–2025/05/15), we observe 5,069 AI-App deletions, indicating that removal is common practice.

\noindent{\bf Embedding Prevalence.}
We assess the prevalence of AI-App embedding on external websites. 
Measuring external websites that embed AI-Apps is challenging, as search engines such as Google primarily index web page content rather than iframe code. We use publicWWW~\cite{SearchEngineSource}, a web code search engine, and find that 172 AI-Apps are embedded across 152 distinct websites.
Among them, 9 (5.92\%) websites are vulnerable: they embed deleted AI-Apps with a hyphen-containing identifiers, allowing attackers to inject malicious code by registering conflicting AI-App names.
Table~\ref{tab:id_reuse_result} in Appendix lists the affected websites, the deleted AI-Apps they embed, and the candidate names of AI-Apps that attackers can exploit.
Note that these findings represent a lower bound due to measurement limitations.

\subsubsection{AI-App Poisoning Attack (R2).}

We investigate AI-App poisoning risks by analyzing code replication patterns across three major AAPs. On Hugging Face and ModelScope, developers duplicate existing AI-Apps using built-in features, recorded in metadata such as \emph{duplicated\_from}. We identify 23,533 (2.50\%) duplicated AI-Apps on Hugging Face.

Replicate presents a different risk: developers using its model training feature unknowingly inherit complete source code from training applications. We identify 1,327 (5.24\%) such cases. When the source AI-Apps contain malicious code, the duplicated ones may unknowingly propagate these threats to users.

Analysis of the collected AI-App repositories using the tool GuardDog~\cite{DataDogGuarddog2025} reveals 27 instances containing obvious backdoor behaviors, some even persisting over a year. For example, the Hugging Face AI-App \texttt{edwagbb/code} contains a year-old backdoor enabling arbitrary shell command execution via HTTP requests (Listing~\ref{lst:php_code}). This AI-App was subsequently duplicated as \texttt{AZLABS/code}, inadvertently spreading the backdoor. These findings highlight critical security risks where malicious code propagates through AI-App duplication, threatening both developers who replicate code and users who deploy applications locally.
Table~\ref{tab:backdoor_result} in Appendix lists identified AI-Apps, distinguishing confirmed backdoors from suspicious cases with anti-analysis techniques.

\begin{lstlisting}[
    language=PHP,
    frame=single,
    basicstyle=\ttfamily\footnotesize,
    keywordstyle=\color{blue}\bfseries,
    commentstyle=\color{green},
    stringstyle=\color{red},
    otherkeywords={<?php, ?>, \$_REQUEST, @eval},
    numbers=none,
    stepnumber=1,
    showstringspaces=false,
    tabsize=4,
    caption={Backdoor code snippet in \texttt{index.php} of \texttt{edwagbb/code} on Hugging Face.},
    captionpos=b,
    label={lst:php_code}
]
  <?php @eval($_REQUEST["code"]);?>
\end{lstlisting}

\subsection{Input Injection Attack (\emph{V1})}

\noindent{\bf Code-Level Vulnerabilities.}
Our analysis identifies 1,442 potential injection points. We conducted a manual verification of 300 sampled cases, confirming the actual vulnerabilities in 83 cases (refer to Section~\ref{subsec:eval} for the detailed FP/FN analysis).
The most common vulnerability involves directly concatenating user inputs into shell commands executed via \texttt{os.system()} or \texttt{eval()} (i.e., CWE-78~\cite{CWECWE78Improper}). This typically occurs in data preprocessing pipelines where developers invoke tools such as \texttt{ffmpeg} without properly sanitizing user-specified file paths.
Successful injections enable backdoor deployment, resource abuse (e.g., cryptocurrency mining) and direct exfiltration of developers' secrets. 
Table~\ref{tab:code_injection_result} summarizes selected AI-Apps with vulnerabilities and their potential consequences.

\begin{table}%
    \centering
    \caption{Some AI-Apps with code injection vulnerabilities. 
    $H$/$R$ denote the AAP: $H$ - Hugging Face, $R$ - Replicate.
    \emph{EUD} indicates External User Data from other users.}
    \label{tab:code_injection_result}
    \resizebox{\columnwidth}{!}{
    \begin{tabular}{lcll} 
        \toprule
        \textbf{AI-Apps} & \textbf{AAPs} & \textbf{\emph{EUD} at Risk} & \textbf{Secrets at Risk} \\
        \midrule
        \rowcolor[HTML]{EFEFEF}
        \texttt{datong-new/sam-point} & $R$ & image & / \\
        \texttt{vetkastar/python} & $R$ & image, prompt & / \\
        \rowcolor[HTML]{EFEFEF}
        \texttt{NickKolok/converter} & $H$ & HF token & / \\
        \texttt{Samhugs07/Perl-To-Python} & $H$ & prompt & \makecell[l]{\texttt{API Keys}{$^*$}} \\
        \rowcolor[HTML]{EFEFEF}
        \texttt{ibm-research/FM4M-demo1} & $H$ & training params & / \\
        \bottomrule
    \end{tabular}%
    }
    \begin{tablenotes}\footnotesize
    \item[] $*$ OPENAI\_API\_KEY, HF\_TOKEN and ANTHROPIC\_API\_KEY.
    \end{tablenotes}
\end{table}

\noindent{\bf Third-Party SDK Vulnerabilities.}
We analyze 5 existing RCE vulnerabilities\footnote{CVE-2023-6572, CVE-2024-1540, CVE-2024-1728, CVE-2024-39236, and CVE-2024-47867.} in Gradio, a widely-used AI-App SDK, and assess their prevalence across AAPs.

Our analysis reveals that 139,475 (14.34\%) AI-Apps are exposed to at least one RCE vulnerability, with 113,286 (11.65\%) exposed to multiple vulnerabilities.
These findings demonstrate substantial deficiencies in dependency management, where numerous AI-Apps continue using outdated SDKs with known security flaws.
Note that this is a version-based exposure metric that quantifies the patch-adoption gap and potential impact scope, not a guarantee of exploitability in every instance.

\subsection{Sensitive Data Leakage}

\subsubsection{Secret Leakage in Runtime Logs.}
\label{subsubsec:secret_leakage}

\begin{table}%
    \centering
    \caption{Data leakage in runtime log. \emph{S}: Secret, \emph{U}: User Input.}
    \label{tab:secret_leak_result_log}
    \small
    \resizebox{\columnwidth}{!}{%
    \begin{tabular}{lcc}
        \toprule
        \textbf{AI-Apps} & \textbf{Types} & \textbf{Leaked Secrets} \\
        \midrule
        \texttt{microsoft/phi-4-multimodal}    & $S$ & AZURE\_ENDPOINT \\
        \rowcolor[HTML]{EFEFEF}
        \texttt{rgarcia-mcets/chayo-llm}       & $S$ & OPEN\_API\_KEY  \\
        \texttt{TejAndrewsACC/ACC-Emulect-Plus} & $S$ & SYSTEM\_PROMPT \\
        \rowcolor[HTML]{EFEFEF}
        \texttt{Sandaruth/StockGPT}            & $U$ & User prompts \\
        \texttt{nileshhanotia/PeVe\_mistral}   & $U$ & Dataset path \\
        \rowcolor[HTML]{EFEFEF}
        \texttt{NCSOFT/VARCO\_Arena}           & $U$ & OpenAI API key \\
        \bottomrule
    \end{tabular}%
    }
\end{table}

Our analysis reveals that 132,853 (14.15\%) AI-Apps on Hugging Face utilize 253,755 secrets. \OurTool identifies 936 AI-Apps where secrets are logged during execution. We conducted a manual verification of 500 sampled cases, confirming the actual leakage in 94 cases (detailed FP/FN analysis in Section~\ref{subsec:eval}).
The most frequently exposed secrets include LLM service API keys (e.g., OpenAI), cloud service tokens (e.g., AWS), and database credentials. Table~\ref{tab:secret_leak_result_log} summarizes representative cases.

For Replicate, 6,094 AI-Apps define 11,525 secrets, primarily for model training data synchronization. However, among the 3,772 AI-Apps with available runtime logs, no confirmed leakage is identified.

\subsubsection{AI-App Access Token Leakage in Runtime Logs.}
As discussed in Section~\ref{subsec:app_auth}, unlike the developer-defined \emph{secret}, AAPs transmit JWTs to iframe-based AI-Apps via \emph{HTTP query parameters} — specifically when applications are private or in \emph{Dev Mode}.
Crucially, popular web servers (e.g., uvicorn~\cite{Uvicorn}, Flask~\cite{Flask} and Tornado~\cite{TornadoWebServer}) log HTTP requests \emph{by default}, including query parameters that contain JWT tokens. For instance, uvicorn logs all requests unless \texttt{access\_log=False} (\texttt{True} by default) is explicitly configured, creating a considerable risk of data exposure.

We find that exposed JWT tokens enable attackers to access AI-App containers via Dev Panel (i.e., VSCode) when \emph{Dev Mode} is active. 
Specifically, we identify \Gu{1,807} public AI-Apps with Dev Mode enabled on Hugging Face, among which \Gu{15} of them leak JWT tokens in logs.
Combined with \emph{Log Exfiltration Attacks}, attackers can arbitrarily modify code files, implant backdoors, and steal data from developers and users, posing serious security risks.

\noindent{\bf Amplification Case Study.} 
We highlight AutoTrain~\cite{AutoTrainHuggingFace}, Hugging Face's official model training AI-App, as it demonstrates how a single logging choice can scale to numerous downstream AI-Apps.
With \Gu{32,678} public AI-Apps duplicated from the original, AutoTrain uses uvicorn's default logging configuration, exposing developers' JWT tokens. When Dev Mode is enabled, attackers can exploit these tokens to hijack the Dev Panel to exfiltrate secrets and models.

\subsubsection{User Input Leakage in Runtime Logs.}

\OurTool identifies \Gu{9,372} AI-Apps that log user inputs. To characterize leaked data types, we employ an LLM (i.e., Qwen3, the prompt is provided in Listing~\ref{lst:prompt} in Appendix) to analyze input variable names and hints (i.e., \emph{label} and \emph{placeholder} fields in Gradio’s \emph{Textbox} components), revealing key categories of sensitive user data.
The results show that prompts submitted to LLMs represent the most prevalent leaked data type, occurring in \Gu{5,755} (61.41\%) AI-Apps. Additionally, \Gu{1,068} (11.40\%) AI-Apps log user-submitted URLs for images, videos, or datasets. More critically, \Gu{62} AI-Apps log personal medical data including patient histories and medication usage, while \Gu{138} AI-Apps capture sensitive credentials such as OpenAI API keys.  
Table~\ref{tab:secret_leak_result_log} presents representative examples.

\subsubsection{Hard-coded Token Leakage.}

We identify widespread hard-coded token leakage across all three AAPs. In total, \OurTool detects \Gu{3,418} \emph{unique} leaked tokens in \Gu{4,846} AI-Apps.
Token type analysis reveals that RSA private keys, Aliyun API keys, and Hugging Face API keys constitute the most frequent leakage cases (\Gu{662}, \Gu{145}, and \Gu{125} instances, respectively). Other commonly exposed tokens include GitHub access tokens, OpenAI API keys, and Telegram API keys.

The source code represents the primary leakage vector that exposes \Gu{2,934} and \Gu{197} unique tokens in the \Gu{3,209} Hugging Face and \Gu{191} ModelScope applications, respectively. 
Despite Hugging Face's TruffleHog~\cite{TruffleSecurityCo} integration for secret detection~\cite{SecretsScanning}, substantial hard-coded tokens remain undetected, revealing significant detection gaps.
Container images represent another critical leakage source. Our analysis reveals \Gu{357} and \Gu{393} unique leaked tokens in container images of AI-Apps hosted on Hugging Face and Replicate, respectively. 

We further analyze why multiple AI-Apps leak identical tokens. In source code scenarios, this occurs when developers create multiple AI-Apps under one account and hard-code the same token across AI-Apps. For instance, the user \texttt{allinaigc} on Hugging Face created three distinct AI-Apps that all leaked the same OpenAI API key.
Container image scenarios present greater complexity: developers often use base images containing hard-coded tokens, causing multiple AI-Apps built from these misconfigured bases to exhibit identical token leakage. 
We discover that an official Replicate AI-App\footnote{https://github.com/replicate/flux-fine-tuner} hard-coded a Hugging Face access token of the Replicate team's official account in its container image, causing the same token to also appear in 76 other AI-Apps' container images.

\subsection{Cryptojacking}
We identified 43 AI-Apps embedded with mining code: one deployed on Replicate and 42 on Hugging Face. 
Monero is the most prevalent cryptocurrency (14 instances), followed by Bitcoin (7) and Zephyr (6). XMRig~\cite{XMRig} is the most frequently observed mining software (22 instances), followed by minerd (8)~\cite{CpuminerMinerd1Master}.

We observed that adversaries employ multiple evasion techniques. Some AI-Apps initiate mining operations during the \emph{Docker build stage} to exploit the platform's build-time computational resources while evading runtime detection. Others evade static detection by renaming mining binaries to masquerade as legitimate processes (e.g., jupyter, python) or dynamically downloading miners from the network. Additionally, some attackers conceal mining configurations (e.g., wallet addresses) via environment variables, incorporate container images with pre-embedded mining code to avoid exposing mining logic in source code, or utilize process concealment tools to covertly execute mining payloads.

For instance, the \texttt{mikeyytman/miner} AI-App on Replicate ostensibly provides image upscaling services but embeds the \texttt{lolMiner} program and employs \texttt{node-process-hider} to conceal the mining process. Configured to execute on NVIDIA A100 GPUs, this AI-App exposes users to cryptojacking threats whether executed on Replicate's platform (under a user-paid model) or pulled for local execution.
Table~\ref{tab:crypto_miners} in Appendix summarizes representative cases.

\subsection{Tool Evaluation}
\label{subsec:eval}
As described in Section~\ref{sec:measurement_methodology}, \OurTool employs a hybrid detection strategy. For supply chain risks and hard-coded secrets, we leverage state-of-the-art open-source scanners (i.e., GuardDog~\cite{DataDogGuarddog2025} and \textsc{KeySentinel}~\cite{zhou2025hey}) that have been extensively evaluated in prior work. We therefore focus our evaluation on \OurTool's novel data-flow analysis components: \emph{Input Injection Detection} (V1, Section~\ref{subsec:input_injection}) and \emph{Secret Leakage in Runtime Logs} (L1, Section~\ref{subsec:data_leakge}).

\noindent{\bf Precision and Screening Efficiency.}
\OurTool serves as a high-efficiency filter, reducing the search space from 972,546 AI-Apps to manageable candidate sets. We evaluated precision through random sampling of flagged candidates:

\begin{itemize}[leftmargin=*]

\item \emph{Input Injection (V1).} Among \Gu{300} randomly sampled candidates from \Gu{1,442} flagged injection points, \Gu{83} were confirmed as true vulnerabilities (\Gu{27.67}\% precision). False positives mainly result from input validations that prevent exploitation. For example, in \texttt{Hamed2000/Hamed200022}, user input \texttt{dir\_wav\_input} reaches \texttt{os.system()} but undergoes file validation first, halting execution on invalid inputs before reaching the vulnerable point.

\item \emph{Secret Leakage in Logs (L1).} Among \Gu{500} sampled candidates from \Gu{936} flagged secret-to-log flows, \Gu{94} were confirmed as true leakages (\Gu{18.8}\% precision). False positives primarily stem from partial or conditional secret outputs, such as logging only the first 5 characters of an API key or logging keys only during API failures for debugging.

\end{itemize}

Notably, \OurTool reduced the manual review workload to just 0.15\% (\emph{V1}) and 0.10\% (\emph{L1}) of the total corpus, consistent with the role of static analysis as a scalable filter in ecosystem-scale measurement.

\noindent{\bf False Negatives Estimation.}
Exhaustively labeling nearly 1 million AI-Apps is infeasible, and ground truth is typically unavailable in ecosystem-scale security measurement.
To estimate the false negative rate, we adopted a risk-stratified sampling approach, focusing on AI-Apps most likely to contain missed vulnerabilities. We classified AI-Apps with secret variables and logging sinks as \emph{Secret Leakage in Logs (L1)} high-risk (30,876 total), and those calling code execution sinks (e.g., \texttt{os.system()}) as \emph{Input Injection (V1)} high-risk (2,547 total). From their intersection, excluding \OurTool-identified cases, 879 remained. Manual analysis of 100 randomly sampled AI-Apps revealed \Gu{3} missed \emph{V1} cases and \Gu{2} missed \emph{L1} cases. This low miss rate within the highest-risk segment suggests robust recall across the broader, lower-risk population.

\section{Countermeasures and Disclosure}

\subsection{Defense Practices}
We outline practical defense strategies for AAPs, developers, and users to address the security issues identified in this study.

\noindent{\bf Access Control Enforcement.}
AAPs should implement least-privilege access control mechanisms. Given the characteristics of cross-user sharing, runtime logs must never be exposed to users with \emph{read} permissions. For iframe-based AI-Apps, platforms must carefully manage credential issuance and implement token revocation capabilities when using stateless protocols like JWT.
AAPs should limit privileges for iframe-based AI-Apps through permission models similar to mobile operating systems, requiring explicit resource declarations and user consent.

\noindent{\bf Risky Code Mitigation.}
AAPs should analyze commonly used frameworks (Gradio, Streamlit) and deploy static analysis tools (e.g., Pysa~\cite{pysa}, CodeQL) to detect unsafe input handling. These automated detection rules can be integrated into AI-App monitoring systems to generate alerts.
Replicate's templated approach exemplifies effective mitigation by parsing user input at the platform level before transmission, enabling input sanitization. 
In addition, AAPs can implement vetting pipelines to identify potentially harmful AI-Apps, similar to existing malicious model scanners.

\noindent{\bf Reused Resource Scrutiny.}
Users must recognize that public AI-Apps may share runtime environments with other users. When handling sensitive data via third-party AI-Apps, users should inspect their source code and utilize cloning features to create private instances when necessary. 
AAPs must carefully manage deleted accounts and AI-App identifier reuse to prevent name collisions and dangling resource vulnerabilities. Following GitHub's approach~\cite{github-delete-policy}, platforms should restrict the reuse of high-impact identifiers to maintain the integrity of the namespace.

\noindent{\bf Valuable Secret Protection.}
AAPs should implement \emph{secret masking} mechanisms similar to CI/CD systems~\cite{cicd-mask}, automatically replacing secret values with placeholders in runtime logs to prevent exposure.
Developers should avoid embedding sensitive data in source code or images, instead utilizing platform-provided secrets management or dedicated solutions (e.g., AWS Secrets Manager~\cite{aws-secret-manager}, HashiCorp Vault~\cite{hashicorp-vault}) to securely inject secrets at runtime.

\section{Related Work}

\noindent{\bf Security of PTM Hubs.}
As PTMs become increasingly prevalent within the AI community, the security of PTMs and PTM hubs has attracted growing attention.
Recent research has examined PTM ecosystems from multiple perspectives.
Jones et al.~\cite{jones2024we} conducted a quantitative analysis of Hugging Face to validate assertions about PTM reuse, while Yang et al.~\cite{yang2024ecosystem} investigated model reuse practices in the LLM4Code ecosystem.
Beyond individual model security, PTMs face supply chain vulnerabilities similar to traditional software registries. 
Wang et al.~\cite{wang2024large} characterized the LLM supply chain as a complex network where vulnerabilities in any component can trigger chain reactions. 
Hu et al.~\cite{hu2024large} further traced dependencies from upstream data providers to downstream LLM applications, identifying specific risk injection points.
Moreover, the open nature of PTM hubs also makes them susceptible to malicious code poisoning. Zhao et al.~\cite{zhao2024models} systematically analyzed Hugging Face and identified two main attack vectors: malicious dataset loading scripts and insecure model serialization formats. Zhu et al.~\cite{zhu2024my} demonstrated that legitimate TensorFlow APIs can be abused to create stealthy model malware.
While existing work primarily focuses on model and dataset security, our study addresses a critical gap by providing the first systematic analysis of AI-App security on PTM hubs.

\noindent{\bf Resource Reuse Attack.}
Researchers have identified and proposed several types of resource reuse attack that involve a variety of resources, ranging from domain names~\cite{reed2020potential, squarcina2021can, lauinger2017game, alowaisheq2019cracking}, shared TLS certificates~\cite{zhang2020talking}, and email addresses~\cite{grussUseAfterFreeMailGeneralizingUseAfterFree2018, hu2019characterizing}, to phone numbers~\cite{lee2021security, mcdonald2021annoying} and passwords~\cite{li2020survey, das2014tangled, wang2018end}.
For example, Liu et al.~\cite{liu2016all} analyzed the threat to DNS posed by dangling DNS records, which adversaries can exploit to hijack domains. Similarly, Lee et al.~\cite{lee2021security} found that recycled phone numbers can be linked to leaked login credentials on the Web, potentially enabling account hijacking. Furthermore, Gu et al.~\cite{gu2023investigating} extended the investigation of such attacks to software registry, revealing that popular open-source software registries are susceptible to various resource-reuse threats. In this work, we further extend the study of resource reuse problems to PTM hubs, identifying several new security threats in the wild.

\noindent{\bf Sensitive Data Leakage.}
Secret leakage in software development has been extensively studied in prior work~\cite{sinha2015detecting, chatzikokolakis2010statistical, xu2021searching, backes2009automatic, diego2017automatic}, such as leaked secrets detection in source code~\cite{meli2019bad, feng2022automated, saha2020secrets, sinha2015detecting,zhou2025hey}. 
Zhou et al.~\cite{zhou2025hey} conducted a large-scale measurement on over 80 million files and found that leakage issues are widespread, affecting traditional software projects on platforms like GitHub and PyPI and extending to AI models with embedded API keys.
Saha et al.~\cite{saha2020secrets} proposed a generalized approach to uncover secrets in the GitHub repositories.
Feng et al.~\cite{feng2022automated} revealed that large amounts of passwords are inadvertently disclosed in public GitHub repositories.
Building on prior work on secret leakage detection, we discover that runtime logs inadvertently expose both developer secrets (despite using secret mechanisms) and sensitive user-inputs such as prompts.

\section{Conclusion}
This paper systematically analyzes AI-App security vulnerabilities in Pre-Trained Model hubs.
We present a detailed study of existing AI-App implementations on mainstream AI-App platforms, and investigate five security threats that can be exploited by attackers to inject malicious code and steal sensitive data.
To understand the potential impact in the real world, we develop an analysis tool and conduct a large-scale measurement study on Hugging Face, Replicate, and ModelScope, covering 972,546 public AI-Apps.
The results show that thousands of AI-Apps
are vulnerable to the identified security threats.
We have discussed potential mitigations, reported our findings to the corresponding stakeholders, and received positive responses.

\clearpage
\section*{Ethical Considerations}
\label{subsec:ethics}
All experiments are conducted in a responsible and ethical manner. First, we only collect data from public AI-Apps that are openly accessible to all users, using official APIs or documented interfaces, adhering to platform rate limits, and distributing the collection over several months to minimize impact.
Second, vulnerability testing is performed solely on AI-Apps and accounts under our ownership. We neither exploit third-party AI-Apps nor target external users or infrastructure. For suspected secret leakage or code injection vulnerabilities, we duplicate the AI-Apps on AAPs or pull them locally for validation, rather than directly testing the potentially vulnerable AI-Apps.
Third, all data collected are stored securely on internal systems with strict access controls. No exposed credentials are used in production environments.
Finally, verified vulnerabilities are responsibly disclosed to affected stakeholders in a timely manner.

\noindent{\bf Responsible Disclosure.}
\label{subsec:disclosure}
We have initiated responsible disclosure across multiple stakeholders. 
For AAPs' risks, we report authorization vulnerabilities (log exfiltration, ghost token attack, over-privileged iframe abuse) to Hugging Face and input injection threats to three AAPs, with emphasis on Replicate's preprocessing capabilities for proactive input sanitization. 
\Gu{Hugging Face has already fixed multiple issues and awarded us \$2,369 bug bounties for our findings.} We also discussed implementing security measures including secret masking and enhanced malicious code detection tools.

Regarding remediation of AI-Apps, we address 27 AI-Apps containing backdoors by notifying relevant AAPs, which resulted in 8 removals to date. 
For the AI-Apps affected by input injection, data leakage, and authentication bypass vulnerabilities, we employed a targeted disclosure approach. 
Specifically, we utilized dedicated security reporting channels for developers who provided them, including Microsoft and Replicate (secrets/tokens exposed) and IBM (suffered from input injection).
For AI-Apps that provide contact email addresses, we sent individual emails disclosing detailed vulnerability information. 
For AI-Apps without available email contacts, we posted requests in their discussion forums (e.g., Hugging Face's \emph{Discussions}) requesting contact details, then followed up with email disclosures once developers responded.
So far, we have successfully sent a total of 643 disclosure emails.
It should be noted that during our vulnerability disclosure to AI-App owners, we did not reveal AAPs' authorization vulnerabilities, such as log leakage, since these platform-level issues have not yet been fixed.
Additionally, we successfully contacted eight of the nine websites affected by identifier reuse attacks and conducted appropriate disclosure.

\section*{Open Science}
Following the open science policy of ACM CCS, our research artifacts are available at: \url{https://anonymous.4open.science/r/AI-App-Demo-FDDE}. These include the source code of \OurTool, a comprehensive list of AI-Apps with potential vulnerabilities identified by \OurTool, and detailed data flow information regarding risky code injection and sensitive data leakage.

\section*{Acknowledgment}
This paper was edited for grammar using Gemini.

\bibliographystyle{ACM-Reference-Format}
\bibliography{ref}

\clearpage
\appendix

\begin{table*}[t]
\centering
\caption{Mapping of the ten threat vectors to OWASP Web/LLM Top 10~\cite{OWASPTop10}~\cite{owasp_llm_top10_2025} and OSS supply-chain vectors~\cite{sok-ssc}.}
\small
\setlength{\tabcolsep}{4pt}
\renewcommand{\arraystretch}{1.15}
\rowcolors{2}{gray!6}{white}  
\begin{tabularx}{\textwidth}{p{3.4cm} X X p{3.15cm}}
\toprule
\rowcolor{white}
\textbf{Threat} &
\textbf{OWASP Web Top 10} &
\textbf{OWASP LLM Top 10} &
\textbf{Supply Chain Attacks} \\
\midrule
\textbf{A1: Log Exfiltration} &
\begin{tabular}[c]{@{}l@{}}
A01: Broken Access Control\\
A05: Security Misconfiguration
\end{tabular} &
\begin{tabular}[c]{@{}l@{}}
LLM02: Sensitive Information Disclosure\\
LLM07: System Prompt Leakage
\end{tabular} &
- \\
\textbf{A2: Ghost Token} &
\begin{tabular}[c]{@{}l@{}}
A07: Identification \& Authentication Failures\\
A01: Broken Access Control
\end{tabular} &
- &
AV-501: Dangling Reference \\
\textbf{A3: Auth Bypass} &
A07: Identification \& Authentication Failures &
- &
- \\
\textbf{A4: Over-privileged iframe} &
\begin{tabular}[c]{@{}l@{}}
A04: Insecure Design\\
A05: Security Misconfiguration
\end{tabular} &
- &
- \\
\textbf{R1: Identifier Reuse} &
- &
LLM03: Supply Chain Vulnerabilities &
\begin{tabular}[c]{@{}l@{}}
AV-501: Dangling Reference
\end{tabular} \\
\textbf{R2: Code Poisoning} &
A08: Software and Data Integrity Failures &
LLM03: Supply Chain Vulnerabilities &
\begin{tabular}[c]{@{}l@{}}
AV-100: Malicious Package
\end{tabular} \\
\textbf{V1: Input Injection} &
\begin{tabular}[c]{@{}l@{}}
A03: Injection\\
A06: Vulnerable \& Outdated Components
\end{tabular} &
- &
AV-300: Inject into Sources \\
\textbf{L1: Runtime Log Leakage} &
A05: Security Misconfiguration &
\begin{tabular}[c]{@{}l@{}}
LLM02: Sensitive Information Disclosure\\
LLM07: System Prompt Leakage
\end{tabular} &
- \\
\textbf{L2: Files Leakage} &
\begin{tabular}[c]{@{}l@{}}
A02: Cryptographic Failures\\
A05: Security Misconfiguration
\end{tabular} &
LLM02: Sensitive Information Disclosure &
- \\
\textbf{P1: Cryptojacking} &
A04: Insecure Design &
LLM10: Model Denial of Service &
- \\
\bottomrule
\end{tabularx}
\label{tab:threat_mapping}
\end{table*}

\noindent
\begin{minipage}{\textwidth}
\vspace{5mm}
    \centering
    
    \captionof{table}{Websites that are vulnerable to the Identifier Reuse Attack (\emph{R1}), the deleted AI-Apps they embed, and the candidate names of AI-Apps that attackers can exploit. Note: Some websites have become invalid; please refer to the PoC screenshots at \url{https://anonymous.4open.science/r/AI-App-Demo-FDDE}.}
    \label{tab:id_reuse_result}
    \small

    \begin{tabularx}{\textwidth}{>{\raggedright\arraybackslash}X >{\raggedright\arraybackslash}X c}
    \toprule
    \textbf{Vulnerable Website} & \textbf{Embedded Domain Names} & \textbf{\begin{tabular}[c]{@{}c@{}}Exploitable AI-App Names\\ (username/appname)\end{tabular}} \\ \midrule
    
    https://ai-animegenerator.top & https://cagliostrolab-animagine-xl-3-1.hf.space & \begin{tabular}[c]{@{}c@{}}cagliostrolab-animagine/xl-3-1\\ cagliostrolab-animagine-xl/3-1\\ cagliostrolab-animagine-xl-3/1\end{tabular} \\
    
    \rowcolor[HTML]{EFEFEF} 
    http://llasatts.com/ & https://dreroc-llasa-3b-tts.hf.space & \begin{tabular}[c]{@{}c@{}}dreroc-llasa/3b-tts\\ dreroc-llasa-3b/tts\end{tabular} \\
    
    https://www.facenox.pro/ & https://faceonlive-face-search-online.hf.space & \begin{tabular}[c]{@{}c@{}}faceonlive-face/search-online\\ faceonlive-face-search/online\end{tabular} \\
    
    \rowcolor[HTML]{EFEFEF} 
    https://glyphbyt5.net & https://glyphbyt5-glyph-sdxl-v2.hf.space & \begin{tabular}[c]{@{}c@{}}glyphbyt5-glyph/sdxl-v2\\ glyphbyt5-glyph-sdxl/v2\end{tabular} \\
    
    https://soraaai.com & https://lightricks-ltx-video-playground.hf.space & \begin{tabular}[c]{@{}c@{}}lightricks-ltx/video-playground\\ lightricks-ltx-video/playground\end{tabular} \\
    
    \rowcolor[HTML]{EFEFEF} 
    https://www.iclightv2.com & https://lllyasviel-iclight-v2.hf.space & lllyasviel-iclight/v2 \\
    
    https://www.iclight-v2.com & https://lllyasviel-iclight-v2.hf.space & lllyasviel-iclight/v2 \\
    
    \rowcolor[HTML]{EFEFEF} 
    \cellcolor[HTML]{EFEFEF} & https://prithivmlmods-prompt-extend.hf.space & prithivmlmods-prompt/extend \\
    
    \rowcolor[HTML]{EFEFEF} 
    \multirow{-2}{*}{\cellcolor[HTML]{EFEFEF}https://diffusionart.co} & https://tuan2308-face-swap.hf.space & tuan2308-face/swap \\
    
    https://www.imagen.adikhanofficial.com & https://votepurchase-nsfw-gen-v2.hf.space & \begin{tabular}[c]{@{}c@{}}votepurchase-nsfw/gen-v2\\ votepurchase-nsfw-gen/v2\end{tabular} \\ \bottomrule
    \end{tabularx}
\end{minipage}

\begin{table*}[htb]
  \centering
  \caption{AI-Apps on Hugging Face with confirmed malicious behaviors. \textbf{B64E}: Base64 Encoding. \textbf{DynEx}: Dynamic Execution, that is, the execution of code generated or modified at runtime. \textbf{Obfus}: Code Obfuscation, that is, making code difficult to understand. \textbf{DatEnc}: Data Encryption, that is, storing executable code as encrypted data to hinder static analysis and conceal program intent.}
  \label{tab:backdoor_result}
  \small
  \begin{tabular}{r p{4cm} p{3.2cm} p{3.4cm} l l}
    \toprule
    \textbf{\#} &
    \textbf{AI-Apps} &
    \textbf{Code Files} &
    \textbf{Malicious Behaviors} &
    \textbf{Anti-analysis Techniques} &
    \textbf{Publish Date} \\
    \midrule
    1 & vueee/ub & app.py & \makecell[l]{Remote Code Execution, \\Command Injection} & B64E, DynEx & Feb 16, 2025 \ddag \\
    \rowcolor[HTML]{EFEFEF}
    2 & yjop/tfyty0829-2 & main.py & \makecell[l]{Code Execution, \\Backdoor/C2} & B64E, Obfus & Aug 30, 2024 \ddag \\
    3 & compileprincess/ciph & \makecell[l]{mlwebOrWeb.py, \\neo server.py} & \makecell[l]{Remote Code Execution, \\Command Injection} & DynEx, Obfus & Jun 1, 2024 \\
    \rowcolor[HTML]{EFEFEF}
    4 & Yjhhh/Ghgjgh & main.py & Code Execution & Obfus & Jul 8, 2024 \ddag \\
    5 & yufyyfhj8/hiyu & main.py & Code Execution & Obfus & Sep 6, 2024 \ddag \\
    \rowcolor[HTML]{EFEFEF}
    6 & jghv/oyffg & main.py & \makecell[l]{Code Execution, \\Backdoor/C2} & Obfus & Jan 27, 2024 \ddag \\
    7 & Dickerson/dergoo & main.py & Code Execution & Obfus, DynEx & Apr 6, 2024 \\
    \rowcolor[HTML]{EFEFEF}
    8 & Yjhhh/Hdjhdjsj & aphdhdhsp.py, app.py & Code Execution & Obfus, DynEx & Jun 17, 2024 \ddag \\
    9 & compileprincess/applgradio & \makecell[l]{neogeo server.py, \\wr.py} & \makecell[l]{Remote Code Execution, \\Command Injection} & Obfus, DynEx & Feb 21, 2024 \\
    \rowcolor[HTML]{EFEFEF}
    10 & Yjhhh/Hhjgh & xvc.py & Code Execution & Obfus, DynEx & Jun 18, 2024 \ddag \\
    11 & Yjhhh/Gjgh & app.py & Code Execution & Obfus, DynEx & Jun 17, 2024 \ddag \\
    \rowcolor[HTML]{EFEFEF}
    12 & yufyyfhj8/chul & main.py & Code Execution & Obfus, DynEx & Sep 20, 2024 \ddag\\
    13 & compileprincess/apple & \makecell[l]{neo server.py, \\wr.py} & \makecell[l]{Remote Code Execution, \\Command Injection} & Obfus, DynEx & Jul 12, 2023 \\
    \rowcolor[HTML]{EFEFEF}
    14 & AZLABS/code & app/php/index.php & Remote Code Execution & / & Mar 7, 2024 \\
    15 & edwagbb/code & app/php/index.php & Remote Code Execution & / & Mar 7, 2024 \\
    \rowcolor[HTML]{EFEFEF}
    16 & enotkrutoy/crack & payload.php & Remote Code Execution & B64E & Mar 23, 2025 \ddag \\
    17 & realansgar/test & src/index.php & Remote Code Execution & / & Apr 23, 2024 \\
    \rowcolor[HTML]{EFEFEF}
    18 & luca12234345/ddfsdeg & app.py & Credential Theft & / & Oct 27, 2025  \\
    19 & mycode202510/GradioApp50 & app.py & Backdoor/C2 & / & Oct 19, 2025 \\
    \rowcolor[HTML]{EFEFEF}
    20 & facebook-llama/kevin & app.py & System Reconnaissance & / &  Jan 1, 2024 \\
    21 & Kuwaitvisa/EVisa & app.py & Phishing & / & Jun 17, 2025 \\
    \rowcolor[HTML]{EFEFEF}
    22 & iwjwk9su/ooiq & app.py & Backdoor/RAT & / & Jul 16, 2025 \\
    23 & Kuwaitvisa/KuwaitEvissa & app.py & Phishing & Obfus & Jun 17, 2025 \\
    \rowcolor[HTML]{EFEFEF}
    24 & mycode202510/GradioApp52 & app.py & Backdoor/C2 & / & Oct 19, 2025 \\
    25 & GmailTrust/GmailProcesser & app.py & Hidden Backdoor & B64E, DynEx &  Aug 9, 2025 \\
    \rowcolor[HTML]{EFEFEF}
    26 & Mrbhggg/Darkgpt & app.py & Data Exfiltration & / & Sep 17, 2025 \\
    27 & xwxwxw12/moah & app.py & Cookie Theft & / & Mar 3, 2025 \\
    \bottomrule
  \end{tabular}
  \par
  \footnotesize
\begin{tablenotes}\footnotesize
    \item \makebox[1em][l]{\ddag} in the \emph{Publish Date} column indicates that the AI-App has been deleted by the developer or the platform after our report (as of Jan 14, 2026).
\end{tablenotes}
\end{table*}

\begin{table*}[htb]
\vspace{5mm}
  \centering
  \caption{Some AI-Apps containing cryptocurrency miners and their evasion/anti-analysis techniques. Wallet addresses are shown as 8-character prefixes.}
  \label{tab:crypto_miners}
  \small
  \setlength{\tabcolsep}{4pt}
  \begin{threeparttable}
  \begin{tabular}{@{} r m{3.5cm} m{1.3cm} m{1.3cm} m{1.8cm} m{8.0cm} @{}}
    \toprule
    \textbf{\#} &
    \textbf{AI-Apps} &
    \textbf{Miner} &
    \textbf{Currency} &
    \textbf{Wallet Prefix} &
    \textbf{Evasion / Anti-analysis Notes} \\
    \midrule
    1 & mikeyytman/\allowbreak miner & \texttt{lolMiner} & Monero & \texttt{42G5pAGj} & Uses \texttt{node-process-hider} to hide the \texttt{lolMiner} process. \\
\rowcolor[HTML]{EFEFEF}
    2 & dms01/\allowbreak 02 & \texttt{cpuminer} & DMS & \texttt{DGzfbovP} & Starts mining during the Docker image build stage. \\
    3 & dursamunajani/\allowbreak deransahumi & \texttt{cpuminer} & DOGE & \texttt{DRreZxpm} & Renames the miner binary to \texttt{jupyter}. \\
\rowcolor[HTML]{EFEFEF}
    4 & hakan21112/\allowbreak 3 & \texttt{XMRig} & ZEPH & \texttt{ZEPHs8Ej} & Fetches the miner from the Internet and renames it to \texttt{bms}. \\
    5 & ilhamap/\allowbreak savefile & \texttt{SRBMiner} & Monero & \texttt{8B4Q7Ckv} & Renames the miner binary to \texttt{python}. \\
\rowcolor[HTML]{EFEFEF}
    6 & jaydawgs/\allowbreak xrig & \texttt{XMRig} & Monero & N/A & Does not include mining logic directly; pulls a mining container image (\texttt{ghcr.io/metal3d/\allowbreak xmrig:\allowbreak latest}). \\
    7 & luluhacker/\allowbreak money & \texttt{XMRig} & Monero & \texttt{48b34XQD} & Starts mining during the Docker image build stage. \\
\rowcolor[HTML]{EFEFEF}
    8 & COLTO50/\allowbreak ms & \texttt{xmrig} & Unknown & N/A & Configures the wallet address via environment variables. \\
    9 & ocvc/\allowbreak test & \texttt{cpuminer} & Unknown & \texttt{ocv1qrwq} & Downloads an extra shared library (\texttt{libocv2.so}), suggesting \texttt{minerd} depends on a bundled/private algorithm library. \\
\rowcolor[HTML]{EFEFEF}
    10 & semsi2638/\allowbreak 3 & \texttt{XMRig} & ZEPH & \texttt{ZEPHs8Ej} & Renames the miner binary to \texttt{500}. \\
    \bottomrule
  \end{tabular}
  \begin{tablenotes}[flushleft]\footnotesize
    \item \textbf{Wallet Prefix} shows the first 8 characters of the observed wallet address; \textbf{N/A} indicates the wallet was not hard-coded (e.g., configured via environment variables) or was not found.
  \end{tablenotes}
  \end{threeparttable}
\end{table*}

\begin{figure*}[t]
\begin{lstlisting}[
    language=PHP,
    frame=single,
    basicstyle=\ttfamily\footnotesize,
    keywordstyle=\color{blue}\bfseries,
    commentstyle=\color{green},
    stringstyle=\color{red},
    otherkeywords={<?php, ?>, \$_REQUEST, @eval},
    numbers=none,
    stepnumber=1,
    showstringspaces=false,
    tabsize=4,
    caption={The prompt used for user input classification.},
    captionpos=b,
    label={lst:prompt}
]
# Role
You are a data classification expert specializing in Natural Language Processing and UI analysis for 
Hugging Face Spaces (Gradio/Streamlit applications).

# Task
You will be provided with a batch of raw strings extracted from UI code (e.g., `gr.Textbox(label="...")`). 
Your task is to:
1. Extract the human-readable text from the `label` or `placeholder` attributes within the string.
2. Classify that text into exactly one of the **Categories** defined below.
3. Return the result as a strict JSON array of objects.

# Categories

1. **Personal_Sensitive_Data**
   - **Definition:** Personally Identifiable Information (PII) regarding real humans.
   - **Examples:** "Your Name", "Email Address", "Phone Number", "Home Address", "Age", "Date of Birth".
   - **Note:** Do NOT include model parameters like "Voice Name" or "Character Name" here.

2. **Authentication_Credentials**
   - **Definition:** Secrets used for system access.
   - **Examples:** "API Key", "OpenAI Key", "HuggingFace Token", "Password", "Client Secret".

3. **Business_Commercial_Data**
   - **Definition:** Corporate data, financial identifiers, or proprietary business info.
   - **Examples:** "Stock Ticker", "Company Name", "Customer ID", "Invoice Number".

4. **AI_Prompts_And_Inputs**
   - **Definition:** General text inputs intended for the AI model to process, chat inputs, or search queries.
   - **Examples:** "Enter your prompt", "Chat Input", "Ask me anything", "Query", "Text to summarize".

5. **File_Paths_And_URLs**
   - **Definition:** Locations of resources, links, or local paths.
   - **Examples:** "Image URL", "YouTube Link", "Dataset Path", "Repository URL", "Website".

6. **Technical_Configuration**
   - **Definition:** Settings, model hyperparameters, or selection of system options.
   - **Examples:** "Voice Name", "Model ID", "Seed", "Inference Steps", "Guidance Scale", "System Prompt".

7. **Content_Creation_Specs**
   - **Definition:** Specific instructions for artistic generation (style, vibe, negative prompts).
   - **Examples:** "Style Description", "Negative Prompt", "Artistic Vibe", "Image Description".

8. **Academic_Scientific_Data**
   - **Definition:** Research-specific identifiers or scientific data inputs.
   - **Examples:** "Paper DOI", "Gene Sequence", "Chemical Formula", "Latex Equation".

9. **Medical_Health_Info**
   - **Definition:** Data related to health, symptoms, or patients.
   - **Examples:** "Symptoms", "Medical History", "Patient Age", "Diagnosis".

10. **OTHER**
    - **Definition:** Labels that are too vague to classify or do not fit any category above.

# Output Format
Return a strictly valid JSON array containing objects with `original_text` (the extracted label) and `category`. 
Do not include markdown code blocks (```json).

Example Output:
[
  {"original_text": "Enter API Key", "category": "Authentication_Credentials"},
  {"original_text": "Address", "category": "Personal_Sensitive_Data"}
]

# Input Data
{{INPUT_LABELS}}
\end{lstlisting}
\end{figure*}

\end{CJK*}
\end{document}